\documentclass[sn-standardnature,iicol]{sn-jnl}
\usepackage{stfloats}

\jyear{2021}%

\theoremstyle{thmstyleone}%
\theoremstyle{thmstyletwo}%

\theoremstyle{thmstylethree}%

\begin{document}

\title[The Reverse Quantum Limit in YbB$_{12}$]{The Reverse Quantum Limit: Implications for Unconventional Quantum Oscillations in YbB$_{12}$}


\author*[1]{\fnm{Christopher A.} \sur{Mizzi}}\email{mizzi@lanl.gov}

\author[1,2,3,4]{\fnm{Satya K.} \sur{Kushwaha}}

\author[5]{\fnm{Priscila F.S.} \sur{Rosa}}

\author[3,4,6]{\fnm{W. Adam} \sur{Phelan}}

\author[6]{\fnm{David C.} \sur{Arellano}}

\author[2,3,7]{\fnm{Lucas A.} \sur{Pressley}}

\author[2,3,7]{\fnm{Tyrel M.} \sur{McQueen}}

\author[1]{\fnm{Mun K.} \sur{Chan}}

\author*[1]{\fnm{Neil} \sur{Harrison}}\email{nharrison@lanl.gov}

\affil[1]{\orgdiv{National High Magnetic Field Laboratory}, \orgname{Los Alamos National Laboratory, \city{Los Alamos}, \state{New Mexico}, \postcode{87545}, \country{USA}}}

\affil[2]{\orgdiv{Institute for Quantum Matter, William H. Miller III Department of Physics and Astronomy}, \orgname{The Johns Hopkins University, \city{Baltimore}, \state{Maryland}, \postcode{21218}, \country{USA}}}

\affil[3]{\orgdiv{Department of Chemistry}, \orgname{The Johns Hopkins University, \city{Baltimore}, \state{Maryland}, \postcode{21218}, \country{USA}}}

\affil[4]{\orgdiv{Platform for the Accelerated Realization, Analysis and Discovery of Interface Materials (PARADIM), Department of Chemistry}, \orgname{The Johns Hopkins University, \city{Baltimore}, \state{Maryland}, \postcode{21218}, \country{USA}}}

\affil[5]{\orgdiv{MPA-Q}, \orgname{Los Alamos National Laboratory, \city{Los Alamos}, \state{New Mexico}, \postcode{87545}, \country{USA}}}

\affil[6]{\orgdiv{MST-16}, \orgname{Los Alamos National Laboratory, \city{Los Alamos}, \state{New Mexico}, \postcode{87545}, \country{USA}}}

\affil[7]{\orgdiv{Department of Materials Science and Engineering}, \orgname{The Johns Hopkins University, \city{Baltimore}, \state{Maryland}, \postcode{21218}, \country{USA}}}


\abstract{Beyond the quantum limit, many-body effects are expected to induce unusual electronic phase transitions. Materials possessing metallic ground states with strong interactions between localized and itinerant electronic states are natural candidates for the realization of such quantum phases. However, the electronic correlations responsible for increasing the likelihood of novel phases simultaneously place the quantum limit beyond the reach of laboratory magnets. Here we propose these difficulties can be surmounted in materials with strong correlations and insulating ground states. Strong correlations in insulators and high magnetic fields conspire to fill Landau levels in the reverse order compared to conventional metals, such that the lowest Landau level is the first observed. Consequently, the quantum limit in strongly correlated insulators is reached in reverse and at fields accessible in laboratories. Quantum oscillations measured at high fields in YbB$_{12}$ are shown to have features consistent with the reverse quantum limit. These include how quantum oscillations move in lock step with the angular evolution of the insulator-metal transition and the field dependence of the quantum oscillation frequency. We argue that close to the insulator-metal transition, the insulating state should be viewed through the lens of a magnetic field-induced electronic instability affecting the lowest Landau level states in the quantum limit.
}




\maketitle

Large magnetic fields have dramatic effects on electron motion~\cite{ashcroft2011}. In metallic systems, these effects are particularly pronounced when the cyclotron energy exceeds the Fermi energy and electrons are confined to their lowest Landau level. The resulting ground state, often termed the quantum limit~\cite{tsui1982}, is highly degenerate and susceptible to instabilities yielding a rich variety of electronic phases including spin- and charge-density waves, fractional quantum Hall states, and excitonic insulators~\cite{fenton1968,yoshioka1981,halperin1987,fradkin1999,khveshchenko2001,jain2007}. The likelihood of such novel phases tends to increase with the strength of electronic correlations, which makes it extremely desirable to explore the quantum limit in metals with strong electronic correlations. $f$-electron metals are an enticing platform for such studies, however their large effective masses place the quantum limit beyond the reach of laboratory magnets~\cite{edwards1996,harrison2011,Kushwaha2019}. Accordingly, quantum limit studies have largely focused on semimetals ($\it{e.g.}$, graphite~\cite{iye1982,zhu2019} and TaAs~\cite{ramshaw2018}) and two-dimensional systems ($\it{e.g.}$, graphene~\cite{geim2007,goerbig2011,moon2012} and GaAs heterostructures~\cite{tsui1982}) where small effective masses and low carrier densities enable experimental access to the quantum limit.

\begin{figure*}[!b]%
\centering
\includegraphics[trim=15 0 100 0, clip, width=1\textwidth]{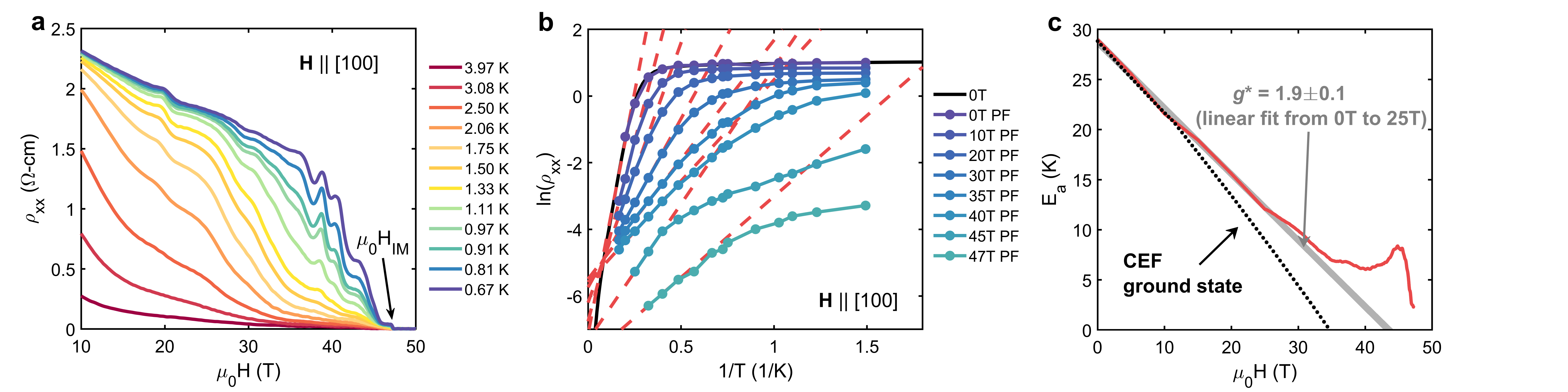}
\caption{(a) YbB$_{12}$ in high magnetic fields exhibits substantial magnetoresistance, quantum oscillations, and a bulk insulator-metal transition ($H_{IM}$). (b) Arrhenius fits (red dashed lines) to the resistivity  measured in pulsed fields (PF) were used to obtain the activation gap at fixed field values. The PF data are in good agreement with zero-field resistivity measurements (black). (c) The activation gap extracted in this manner ($E_a$, red) closes nearly linearly with field up to 25 T, consistent with Zeeman splitting-induced gap closure (grey line). Deviations from linear-in-field gap closure occur above 35 T prior to the insulator-metal transition and do not appear to be consistent with crystal electric field (CEF) level mixing (black, dotted). CEF data taken from Ref.~\cite{Terashima2017}. See Supplementary Information for measurement and analysis details~\cite{SI}.}\label{fig1}
\end{figure*}

In this paper, we propose the quantum limit in systems with strong electronic correlations is most readily reached in reverse, $\it{i.e.}$, from an insulating ground state. This ``reverse quantum limit'' occurs when the Zeeman energy exceeds the cyclotron energy in an insulator and results in reverse Landau level filling. First, we motivate the idea that the first Landau level crossing may correspond to the lowest Landau level by presenting high-field measurements on the Kondo insulator YbB$_{12}$. We establish YbB$_{12}$ as an ideal system to investigate reverse Landau level filling because it has (1) large Zeeman and small cyclotron energies, (2) a small gap (which sets a field scale amenable to laboratory magnets), and (3) reproducible high-field, low-temperature behavior (including unconventional quantum oscillations~\cite{XiangScience,Xiang2021,liu2022}). Then, we demonstrate the angular dependence of the quantum oscillations matches that of the insulator-metal transition field. To explain this observed pinning to the transition field, we formally introduce the reverse quantum limit and show this angular dependence and the field-dependent frequency of the quantum oscillations are manifestations of reverse Landau level filling. Lastly, we show the reverse quantum limit framework captures key aspects of the insulating state quantum oscillations, which suggests the insulating and metallic state quantum oscillations in YbB$_{12}$ share a common origin.

YbB$_{12}$ possesses small gaps of order meV at low temperatures (see SI~\cite{SI}) arising from hybridization between conduction electrons and largely localized $f$-electrons~\cite{Riseborough2000}. When subjected to magnetic fields, YbB$_{12}$ exhibits a large, negative magnetoresistance and undergoes a bulk insulator-metal transition (Fig. \ref{fig1}a). An Arrhenius analysis (Fig. \ref{fig1}b) indicates the insulator-metal transition is driven by field-induced gap closure~\cite{Sugiyama1988}    (Fig. \ref{fig1}c). At first, the activation gap closes linearly with field, followed by notable deviations from linear gap closure above $\sim\!35$T. The linear gap closure is consistent with Zeeman splitting of the conduction and valence band edges corresponding to $g^{*}=$1.9$\pm$0.1~\cite{gDefinition} (Fig. \ref{fig1}c). One possibility is that the high-field behavior may be related to crystal field level mixing~\cite{Moll2017}. However, the predicted crystal field ground state has an effective $g$ factor which is too large to explain our observations~\cite{Terashima2017,Kurihara2021}. It is also possible the non-monotonic behavior is evidence of an excitonic insulator, a possibility that will be explored in greater detail later. Note, the zero field gap, transition field, and effective $g$ factor reported here are consistent with other measurements on high-quality YbB$_{12}$ single crystals \cite{Sugiyama1988, Iga1998, Terashima2017, XiangScience}.

\begin{figure}[!t]%
\centering
\includegraphics[trim=30 0 50 0, clip, width=0.475\textwidth]{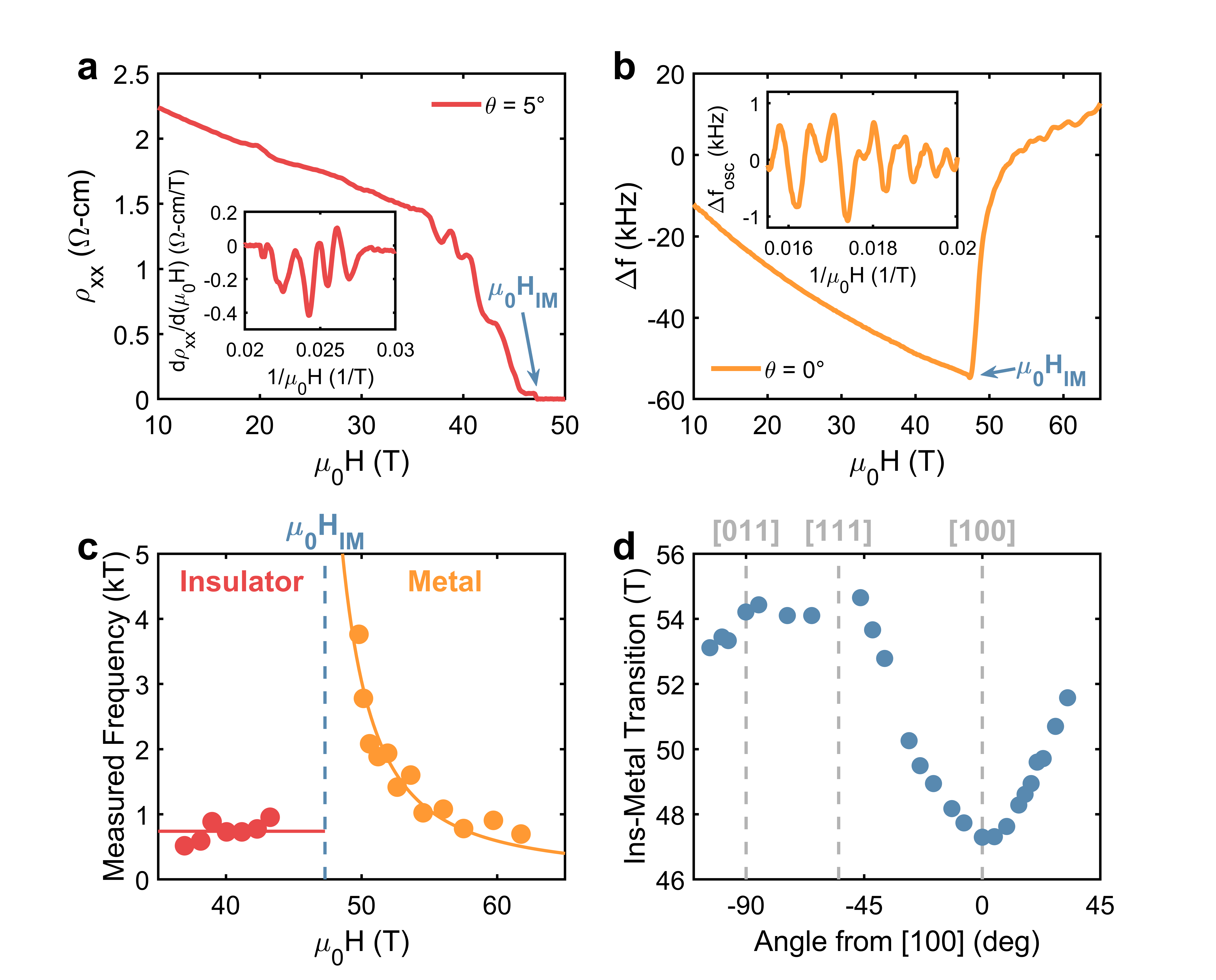}
\caption{(a) Magnetoresistance and (b) contactless resistance showing quantum oscillations in the insulating phase and metallic phase, respectively. Insets show (a) derivative of magnetoresistance and (b) background subtracted contactless resistance as functions of inverse field. (c) Insulating state quantum oscillations have a field-independent frequency of 740$\pm$60 T and metallic state quantum oscillations have a field-dependent frequency ($\bf{H}\parallel$ [100]). Lines are guides for the eye. (d) Angular dependence of the insulator-metal transition field ($H_{IM}$). All measurements were at $\sim$650 mK. Angles correspond to rotations in the [100]-[011] plane referenced to [100].}\label{fig2}
\end{figure}

Magnetoresistance oscillations are present in the insulating state of YbB$_{12}$ for $\mu_0H\gtrapprox35$ T (Fig. \ref{fig1}a and Fig. \ref{fig2}a). The insulating state oscillations are periodic in inverse field with a dominant frequency $\sim$700 T. Frequencies were determined by indexing  maxima and minima of the oscillations and then either finding the slope with respect to inverse field (insulating state) or computing finite differences (metallic state); see Supplementary Information for additional details~\cite{SI}. Temperature dependent amplitudes are consistent with the Lifshitz-Kosevich (LK) form ($m^{*}\approx7.6m{_e}$, see SI~\cite{SI}), which suggests these insulating oscillations are Shubnikov–de Haas (SdH) quantum oscillations~\cite{Shoenberg1984}. SdH quantum oscillations in the field-induced metallic state (Fig. \ref{fig2}b) were observed using tunnel-diode oscillator (TDO) measurements, which are a contactless resistance method~\cite{Coffey2000}. The metallic state quantum oscillations possess a field-dependent frequency (Fig. \ref{fig2}c) and, like the insulating state oscillations, have temperature dependent amplitudes described by the LK equation. Fitting with the LK equation indicates the metallic state effective mass is large ($\sim\!10m_{e}$), increases with field, and has little anisotropy (see SI~\cite{SI}). Importantly, the quantum oscillations in Fig. \ref{fig2} are in good agreement with previous reports in both the insulating \cite{XiangScience,Xiang2022} and metallic states \cite{Xiang2021, liu2022} of YbB$_{12}$. This demonstrates quantum oscillations in high-quality YbB$_{12}$ are a robust and reproducible phenomenon.

\begin{figure*}[!t]%
\centering
\includegraphics[width=0.9\textwidth]{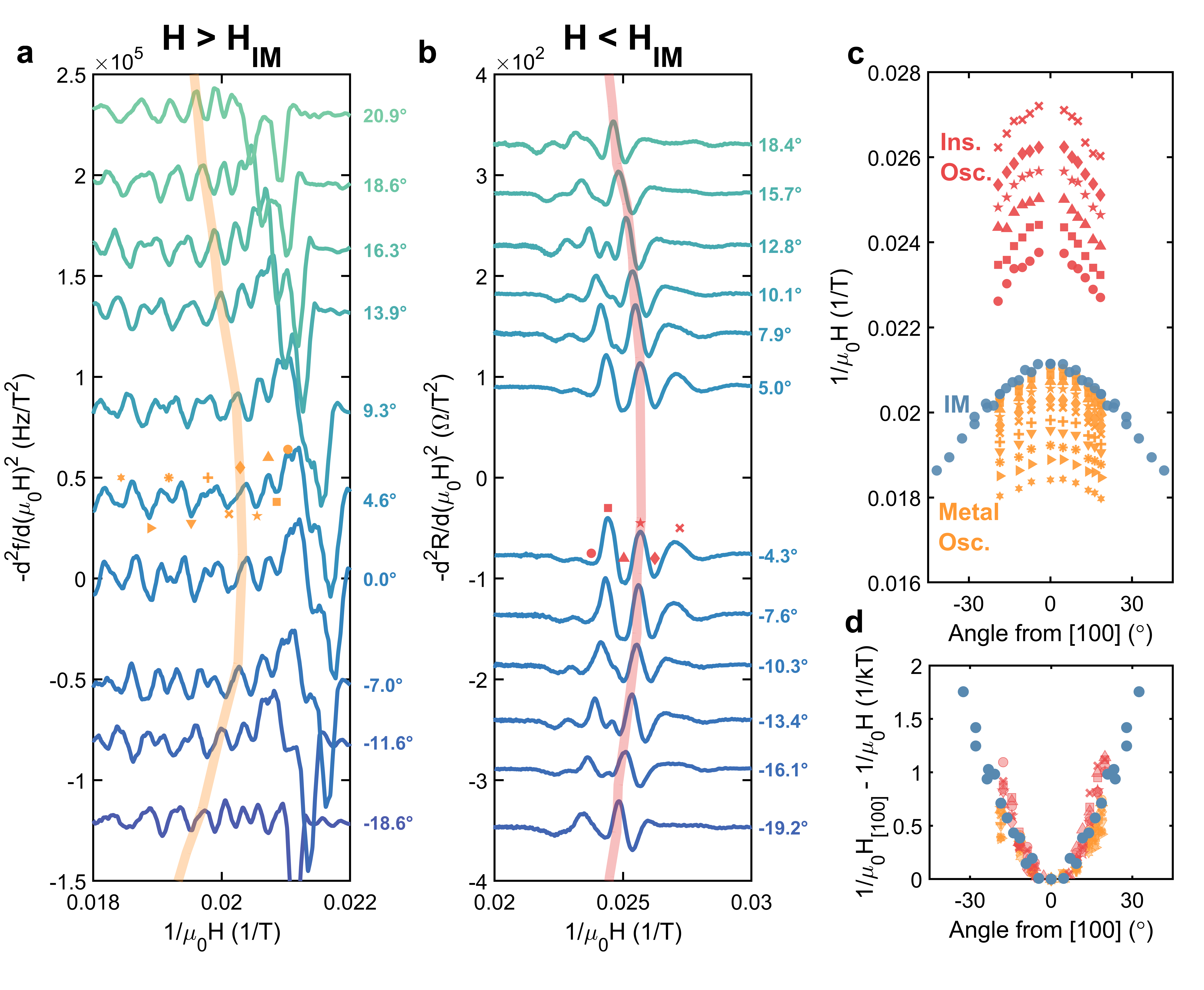}
\caption{Second derivatives of (a) contactless resistance and (b) magnetoresistance as functions of inverse field for fields applied in the [100]-[011] plane (0$^{\circ}$ corresponds to [100]). Data are vertically offset in proportion to angle from [100]. Maxima and minima of oscillations are denoted by symbols and lines are included as guides for the eye. Second derivatives were used to avoid artifacts stemming from background subtraction, and average positions were tracked in cases of split oscillations; see Supplementary Information for details~\cite{SI}. (c) Tracking the angular evolution of individual oscillations (different symbols) in the metallic and insulating states and (d) referencing each to its value when $\bf{H}\parallel$ [100] demonstrates the angular dependence of quantum oscillations follows the insulator-metal transition for a large range of angles and fields in both the (a) metallic and (b) insulating states. All measurements were at $\sim$650 mK.}\label{fig3}
\end{figure*}

The onset of the field-induced metallic state is anisotropic, increasing when the direction of the applied field is rotated away from [100] towards [011] ($H_{IM}^{[011]} / H_{IM}^{[100]}=1.15$, Fig. \ref{fig2}d). We attribute this anisotropy to an anisotropic $g^*$, which is consistent with crystal field predictions~\cite{Terashima2017}. For fields applied in the [100]-[011] plane, the angular dependence of the quantum oscillations measured at base temperature reveals that both the metallic and insulating state quantum oscillations indices follow the angular dependence of the insulator-metal transition; $i.e.$, the oscillations are pinned to the insulator-metal transition. Although there is some evolution in the fine structure of the oscillations, the dominant angular dependence of the oscillations matches the angular variation of the insulator-metal transition over a large angular and field range (Fig. \ref{fig3}). The relationship between the quantum oscillations and the insulator-metal transition is most clearly demonstrated by tracking both phenomena at different angles (Fig. \ref{fig3}c); the angular dependence of each quantum oscillation and the insulator-metal transition collapses to a single curve when referenced to values at $\bf{H}\parallel$ [100] (Fig. \ref{fig3}d). Therefore, the quantum oscillations are pinned to the insulator-metal transition.

\begin{figure}[!b]%
\centering
\includegraphics[trim=20 0 40 20, clip, width=0.45\textwidth]{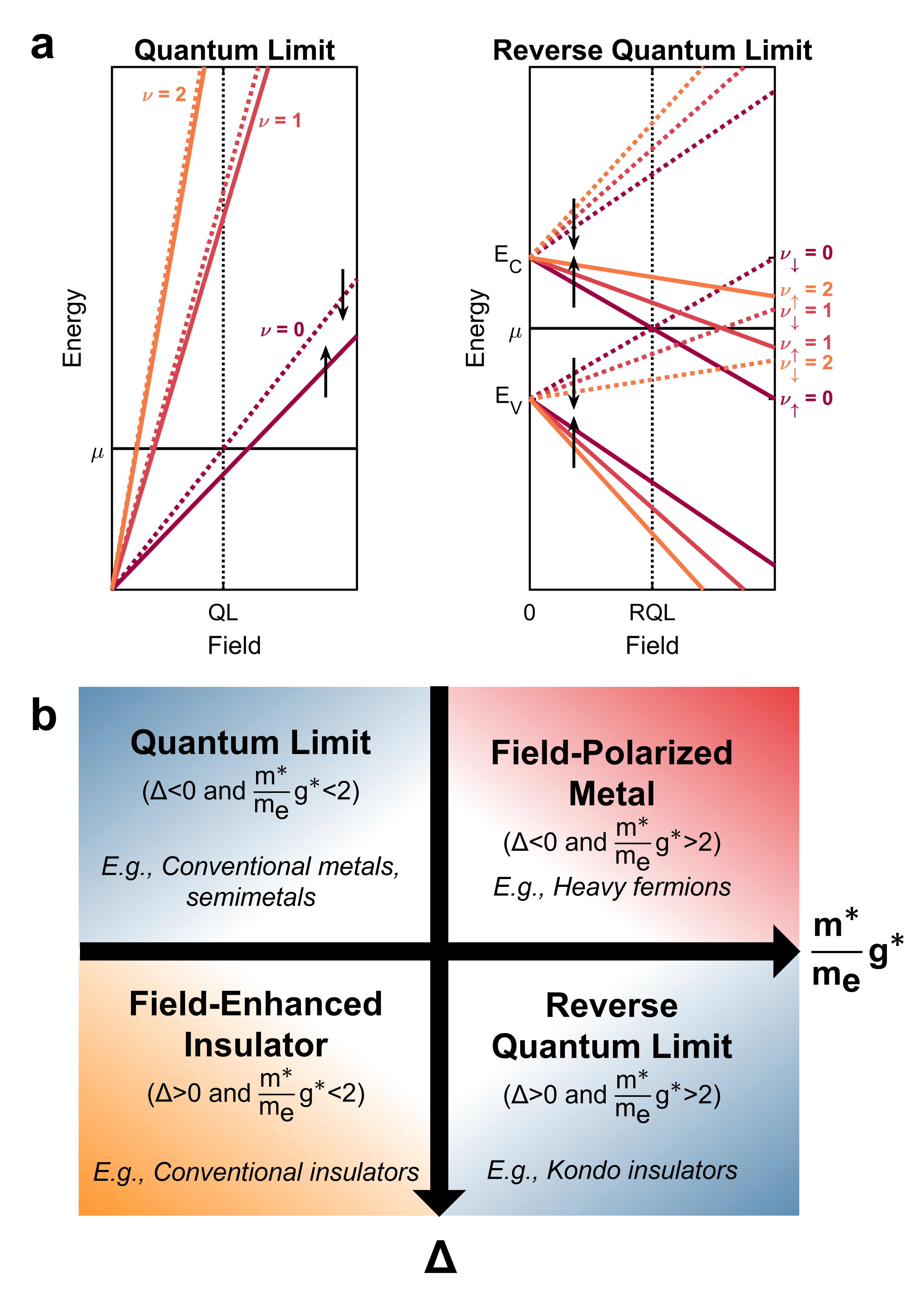}
\caption{(a) Landau level diagrams for the quantum limit (QL) and reverse quantum limit (RQL). A conventional metal reaches the QL when the cyclotron energy exceeds the Fermi energy. An insulator with a Zeeman energy greater than the cyclotron energy is similar, except the Landau levels cross the chemical potential in reverse. (b) High-field electronic phase diagram considering if the electronic structure at the chemical potential is gapped ($\Delta\!>\!0$) or metallic ($\Delta\!<\!0$), and the strength of electronic correlations ($g^{*} \frac{m^*}{m_e}$).}\label{fig0}
\end{figure}

To understand the surprising observation of Landau level indices that are tied to the metal-insulator transition, we contrast the behavior of Landau levels in a conventional metal with those of an insulator and, in doing so, introduce the reverse quantum limit. Electronic states in a conventional metal are quantized by a magnetic field into Landau levels~\cite{ashcroft2011,Shoenberg1984}. When Zeeman splitting is considered, the Landau levels are given by
\begin{equation}
E^{\uparrow, \downarrow} - \mu = \frac{\hbar e}{m^{*}} B \left(\nu+\frac{1}{2}\right) \mp \frac{1}{2} g^{*} \mu_{B} B, \label{eq0}
\end{equation}
where $E^{\uparrow, \downarrow}$ is the energy of the up/down ($-, +$) spin state referenced to the chemical potential $\mu$, $B$ is magnetic field ($B\approx\mu_0H$ for the applied magnetic fields used in this work~\cite{Xiang2021}), $\hbar$ is Plank’s constant, $e$ is electron charge, $m^*$ is effective mass, $\nu$ is the Landau level index, $g^{*}$ is an effective $g$-factor (for pseudospins of 1/2 that is renormalized by interactions), and $\mu_{B}$ is the Bohr magneton. Upon increasing field, Landau levels with decreasing indices sequentially cross the chemical potential. This process continues until the lowest Landau level is reached, corresponding to a conventional metal in the quantum limit (Fig. \ref{fig0}a).

Electron states in insulators are also quantized by magnetic fields, similar to the case described by Eq. (\ref{eq0}). For illustrative purposes, we assume a symmetric band structure with the chemical potential at the center of a zero-field gap ($\Delta$). Note, factors such as disorder will reduce the activation gap ($E_a$, Fig.~\ref{fig1}) from its thermodynamic value ($\Delta$); see Fig.~\ref{fig4} for further clarification. Then, the conduction band edge ($E_C$) in the presence of a magnetic field is given by 
\begin{equation}
E_{C}^{\uparrow, \downarrow} - \mu = \Delta + \frac{\hbar e}{m^{*}} B \left(\nu+\frac{1}{2}\right) \mp \frac{1}{2} g^{*} \mu_{B} B\label{eq1}.
\end{equation}
Landau level crossings ($E_{C}=\mu$)  will only occur if the Zeeman energy exceeds the cyclotron energy, $\it{i.e.}$ if $g^{*} \frac{m^*}{m_e}>2$, where $m_e$ is the bare electron mass (see SI~\cite{SI}). The quantity $g^{*} \frac{m^*}{m_e}$ can be considered a proxy for the strength of electronic correlations. When this condition is met, increasing the field causes Landau levels with increasing indices to sequentially cross the chemical potential such that the lowest Landau level is the first to cross (Fig. \ref{fig0}a). Therefore, the insulator is in the quantum limit at the first Landau level crossing. We call this the reverse quantum limit.

Another way to understand the reverse quantum limit is to consider the reverse quantum limit criterion, $g^{*} \frac{m^*}{m_e}$, in conjunction with the electronic structure at the chemical potential---gapped ($\Delta\!>\!0$) or metallic ($\Delta\!<\!0$). Together, $g^{*} \frac{m^*}{m_e}$ and $\Delta$ provide a means to classify materials according to their behavior in high magnetic fields (Fig. \ref{fig0}b). The quantum limit represents the most familiar case. It is realized in systems with light carriers (and/or small Zeeman energies), which have states at the chemical potential. If such a system were to be gapped it would behave as a field-enhanced insulator, whereas the introduction of stronger electronic correlations would yield a field-polarized metal. The remaining combination of $g^{*} \frac{m^*}{m_e}$ and $\Delta$ describes a system which is the reverse of the quantum limit, $\it{i.e.}$, gapped at the chemical potential with heavy carriers (and/or large Zeeman energies).

Returning to YbB$_{12}$, we find the experimentally determined Zeeman energy ($g^{*}$ = 1.9) and effective mass ($m^*\sim10m_e$) satisfy the reverse quantum limit criterion ($g^{*} \frac{m^*}{m_e}>2$). Furthermore, Eq. (\ref{eq1}) implies Landau level crossings occur when

\begin{equation}
\frac{1}{B_{IM}} - \frac{1}{B_{\nu}} = \frac{\hbar e}{\Delta m^{*}} \nu\label{eq2}
\end{equation}

\noindent where $B_{IM}$ is the insulator-metal transition field and $B_{\nu}$ is the field at which Landau level $\nu$ crosses the chemical potential. Because the right-hand side of Eq. (\ref{eq2}) only weakly depends on angle in the vicinity of the insulator-metal transition (see SI~\cite{SI}), quantum oscillations are expected to be pinned to the insulator-metal transition in the reverse quantum limit, consistent with the data in Fig. \ref{fig3}.

\begin{figure}[!t]%
\centering
\includegraphics[trim=25 0 50 0, clip, width=0.475\textwidth]{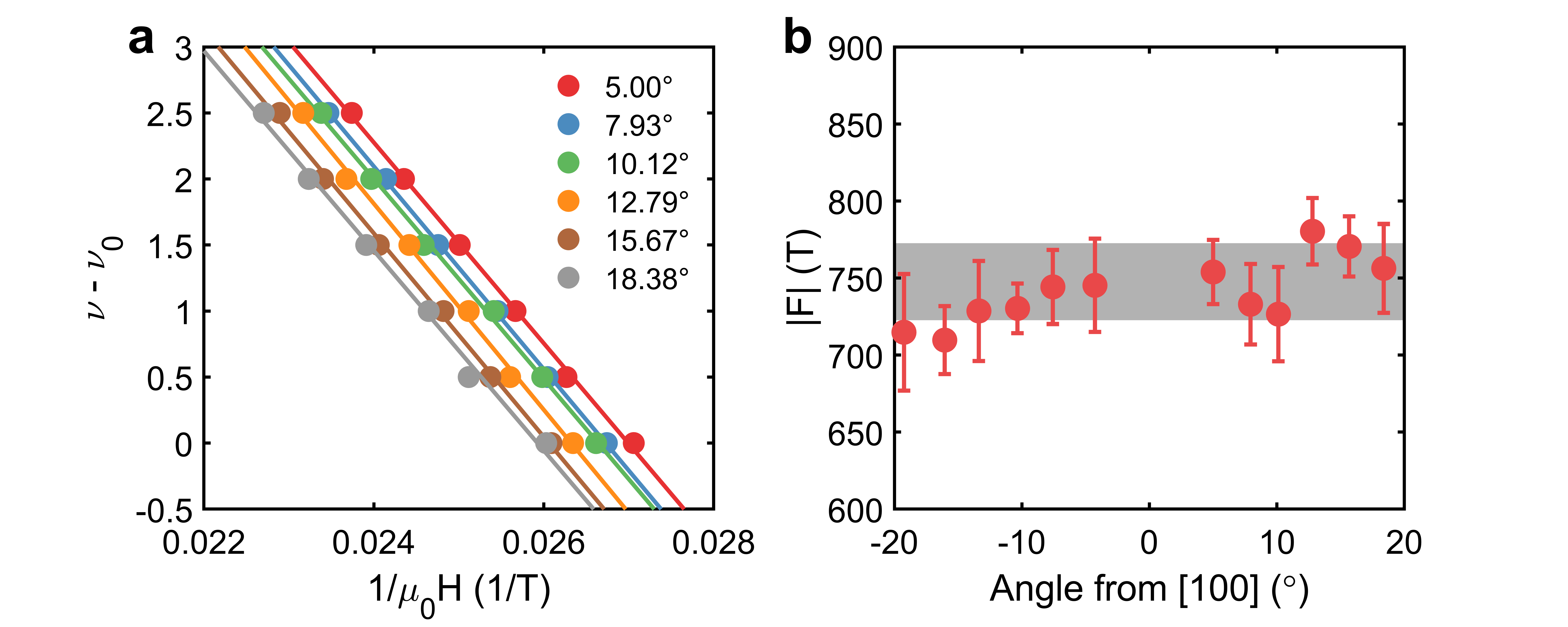}
\caption{(a) Landau level indices ($\nu$) as a function of inverse field in the insulating state for the magnetic field applied along different angles with respect to [100]. Landau levels were indexed by tracking maxima and minima of the oscillations and are reported relative to the first observed oscillation. The indices follow the reverse quantum limit convention, $\it{i.e.}$, lower Landau levels occur at lower fields. Linear fits to the data (solid lines) show nearly identical slopes. (b) The quantum oscillation frequencies corresponding to the slopes in (a) show minimal variation with angle. All data was acquired at $\sim$650 mK. Additional data provided in SI~\cite{SI}.}\label{fig3_5}
\end{figure}

\begin{figure}[!t]
\centering
\includegraphics[trim=240 0 80 0, clip, width=0.475\textwidth]{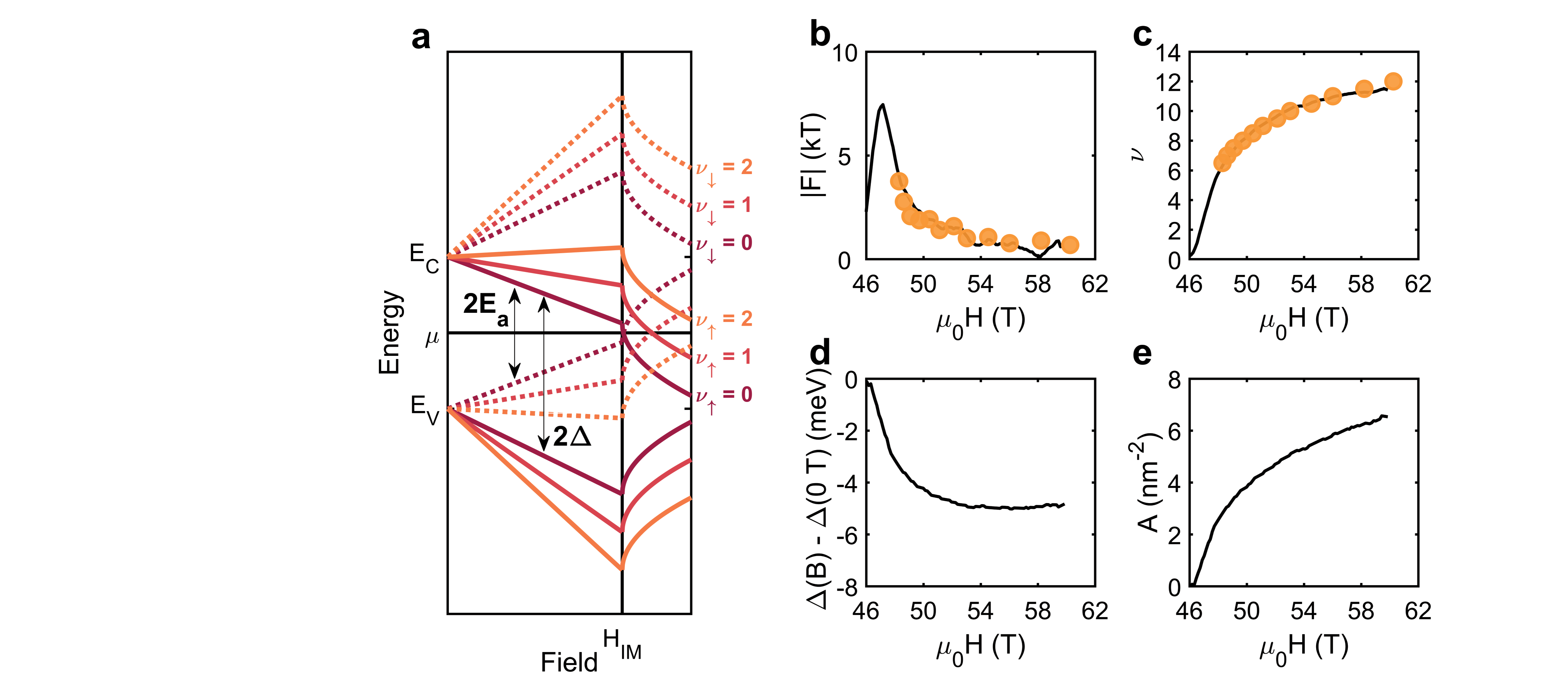}
\caption{(a) Landau level plot considering the combined effects of Zeeman splitting, Landau quantization, and gap reduction at high fields. Above the insulator-metal transition the Landau levels have a non-linear field dependence owing to the non-linear magnetization. (b) Quantum oscillation frequency ($F$) computed assuming the Fermi surface area is proportional to $M^{2/3}$ (black) compared to the measured quantum oscillation frequency in the metallic state (orange). From this relationship it is possible to determine the (c) Landau indices ($\nu$) of the quantum oscillations and the field-dependence of the (d) gap ($\Delta(B) - \Delta(0 T)$) and (e) Fermi surface area ($A$). Magnetization data is taken from Ref.~\cite{Xiang2021}.}\label{fig4}
\end{figure}

Furthermore, Eq.(~\ref{eq2}) corresponds to a quantum oscillation frequency, \begin{equation}
    \lvert F \rvert = \frac{m^{*}}{\hbar e} \Delta, \label{eq2_5}
\end{equation}
\noindent which is proportional to the product of the effective mass and hybridization gap. As both of these quantities exhibit little angular dependence (see SI~\cite{SI}), the reverse quantum limit scenario predicts a quantum oscillation frequency which does not depend on angle. As shown in Fig.~\ref{fig3_5}, the Landau level indices in the insulating state exhibit a linear relationship with inverse field characterized by a slope that varies minimally with angle, consistent with Landau levels filling in reverse.

The field-dependent quantum oscillation frequency (Fig. \ref{fig2}c) also arises naturally in the reverse quantum limit if the hybridization gap varies with field. If $\Delta$ becomes field-dependent in the metallic state, Eq. (\ref{eq1}) implies the quantum oscillation frequency is

\begin{equation}
F(B) = \frac{m^{*}}{\hbar e} \left(B \frac{d\Delta(B)}{dB} - \Delta(B)\right),\label{eq3}
\end{equation}

\noindent where $\Delta(B)$ denotes the field-dependent gap. Thus, the quantum oscillation frequency can exhibit a non-linear field dependence if the gap in the metallic state is a non-linear function of field (see SI~\cite{SI}). Similar considerations explain the slight deviations from the insulator-metal transition at higher angles in the metallic state in Fig. \ref{fig3}d (see SI~\cite{SI}). Note, while $\Delta$ is a constant in the insulating state corresponding to the zero-field gap at the chemical potential, it is also related to the spacing between the zeroth Landau levels of the conduction and valence bands for a given spin at finite field (Fig.~\ref{fig4}a). Therefore, this quantity is well-defined in both the insulating and metallic states.

One possible origin for a non-linear $\Delta(B)$ is a situation in which the magnetic field alters the hybridization between conduction electrons and largely localized $f$-electrons~\cite{edwards1996, Aoki2013}. A scenario in which the hybridization gap is reduced by the applied field is consistent with the appearance of non-linear magnetization in the field-induced metallic state \cite{Sugiyama1988, Terashima2017, Xiang2021} owing to partial $f$-electron polarization: $f$-electron fluctuations are suppressed by $f$-electron polarization at high fields, which disrupts the Kondo-type mechanism responsible for the gap~\cite{Aoki2013}.

To formalize the connection between the gap and extent of $f$-electron polarization, $\it{i.e.}$, the non-linear portion of the magnetization in the high-field metallic state, we assume the Fermi surface area ($A$) is related to the non-linear component of the magnetization ($M$) through

\begin{equation}
A(B) = a \left( \frac{M(B)}{\mu_B} \right)^{2/3}\label{eq4}
\end{equation}

\noindent where $a$ is a constant related to the degeneracy factor and the $2/3$ power arises from assuming ellipsoidal pockets (see SI~\cite{SI}). With this assumption, the measured quantum oscillation frequency is

\begin{equation}
F_{m} = -B^2 \frac{d}{dB}\left[\frac{\hbar}{2 \pi e} \frac{1}{B}  a \left(\frac{M(B)}{\mu_B}  \right)^{2/3}\right].\label{eq5}
\end{equation}

\noindent As shown in Fig. \ref{fig4}, applying Eq.~(\ref{eq5}) yields good agreement between the field dependence of our measured quantum oscillation frequencies in the metallic state and literature values for the magnetization~\cite{Xiang2021} with a single tunable parameter ($a\approx3$). Fixing this $a$ value also gives the field-dependence of the Landau indices and hybridization gap (see SI~\cite{SI}). An example of the non-linear field-dependence of the Landau levels implied by Eq.~(\ref{eq4}) is depicted in Fig. \ref{fig4}a.

The above analysis shows the reverse quantum limit explains the quantum oscillations in the metallic state of YbB$_{12}$: the metallic quantum oscillations track the angular dependence of the insulator-metal transition because they arise from the Landau quantization of bulk bands in an insulator which has a Zeeman energy which exceeds the cyclotron energy, and they exhibit a field-dependent frequency owing to the reduction of the hybridization gap in the high-field metallic state. In other words, the Landau indexing and the insulator-to-(semi)metal transition are inextricably linked in the reverse quantum limit scenario; such a link would not exist were the carriers contributing to metallic conduction not from Landau-quantized states. 

Moreover, the insulating state oscillations exhibit a similar connection to the insulator-metal transition (Fig. \ref{fig3}) which strongly suggests they are a bulk phenomenon and manifestations of the same bulk Landau levels responsible for the metallic state quantum oscillations. Using Eq.~(\ref{eq2_5}) with the measured mass and quantum oscillation frequency in the insulating state yields $\Delta\!\sim\!8$ meV, which is larger than transport gaps ($\it{e.g.}$, Fig. \ref{fig1}), but consistent with optical gaps~\cite{okamura2005}. Since factors such as disorder can reduce the transport gap from its thermodynamic value, our reverse quantum limit interpretation suggests the oscillations in the insulating state are related to the thermodynamic gap.

The reverse quantum limit implies entry into the insulating regime with decreasing magnetic field (and entry into the metallic regime with increasing magnetic field) meets some of the preconditions for the realization of magnetic field-induced electronic instabilities~\cite{fenton1968,yoshioka1981,halperin1987,fradkin1999,khveshchenko2001,jain2007}. One possibility is a transition to an excitonic insulator at high fields prior to the insulator-metal transition in analogy with other quantum limit systems like graphite~\cite{iye1982,zhu2019} and TaAs~\cite{ramshaw2018}, but with much stronger electronic correlations. Such a transition is most likely to occur at the lowest Landau levels, $\it{i.e.}$, close to the insulator-metal transition, and would cause the insulator-metal transition to occur at a higher value of $\nu$ than 0. As our lowest observed Landau level in the metallic state appears to be $\nu\!\approx\!7$ (Fig. \ref{fig4}), this is a possibility. Additionally, an excitonic insulator phase could gap the Landau levels in the (reverse) quantum limit~\cite{yoshioka1981}, perhaps explaining the resistivity oscillations in the insulating phase at high fields. An excitonic transition would also explain the deviations from linear-in-field gap closure for $\mu_{0}H\!\gtrapprox\!35$ T (Fig. \ref{fig1}c) and is consistent with the insulator-metal transition being first order~\cite{Terashima2017,Iga2010}. While we do not observe deviations from LK behavior (predicted for some scenarios with quantum oscillations arising in excitonic insulators~\cite{Knolle2015, Zhang2016, Knolle2017, He2021}), these deviations may only be measurable at lower temperatures than were attainable in these experiments (see SI~\cite{SI}). 

There are currently numerous proposals for the origin of the unconventional, insulating state quantum oscillations in YbB$_{12}$ which invoke concepts such as charge-neutral Fermi surfaces~\cite{Coleman1994, Baskaran2015, Chowdhury2018, Sodemann2018} and non-Hermitian in-gap states~\cite{Shen2018}. Our work indicates a common set of bulk Landau levels drive the insulating and metallic oscillations; however, exactly how these Landau levels yield resistivity oscillations while the bulk remains electrically insulating remains to be addressed. Future work will be needed to understand the consequences of the reverse quantum limit on alternative proposals.

In summary, we have introduced a quantum limit analogue in insulators, the reverse quantum limit, which occurs when the Zeeman energy exceeds the cyclotron energy. The Kondo insulator YbB$_{12}$ is an ideal system to observe this phenomenon because it possesses an electronic structure which satisfies the reverse quantum limit criterion while having an exceptionally small gap which can be closed by fields accessible in the laboratory. Predictions from the reverse quantum limit model are shown to explain how quantum oscillations are pinned to the insulator-metal transition and have a field-dependent frequency in the metallic state. While a detailed mechanism for the insulating state quantum oscillations remains elusive, our work suggests they originate from the same bulk Landau levels as the metallic quantum oscillations, and provides important empirical benchmarks for theoretical descriptions of quantum oscillations in YbB$_{12}$ by demonstrating the central role of the insulator-metal transition.

\bmhead{Acknowledgments}
This work was supported by the Department of Energy (DOE) Basic Energy Sciences (BES) project “Science of 100 Tesla.” The National High Magnetic Field Laboratory is funded by National Science Foundation (NSF) Cooperative Agreements No. DMR-1157490 and No. 1164477, the State of Florida, and DOE. C.A.M. and M.K.C. were supported by the LANL LDRD Program, Project No. 20210320ER. M.K.C. acknowledges support from NSF IR/D program while serving at the National Science Foundation. Any opinion, findings, and conclusions or recommendations expressed in this material are those of the author(s) and do not necessarily reflect the views of the National Science Foundation. S.K.K. acknowledges support of the LANL Directors Postdoctoral Funding LDRD program. This work made use of the synthesis facility of the Platform for the Accelerated Realization, Analysis, and Discovery of Interface Materials (PARADIM), which is supported by the NSF under Cooperative Agreement No. DMR-2039380. Work at Johns Hopkins University was supported by the Institute for Quantum Matter, an Energy Frontier Research Center funded by DOE, Office of Science, BES under Award No. DE-SC0019331. Additional data related to the crystal growth is available at \url{https://doi.org/10.34863/xxxxx}. The authors thank Joe D. Thompson for performing the susceptibility measurements in the Supplementary Information and Boris Maiorov for helpful discussions.

\bibliography{sn-bibliography}

\begin{thebibliography}{10}
\expandafter\ifx\csname url\endcsname\relax
  \def\url#1{\burl{#1}}\fi
\expandafter\ifx\csname urlprefix\endcsname\relax\def\urlprefix{URL }\fi
\providecommand{\bibinfo}[2]{#2}
\providecommand{\eprint}[2][]{\url{#2}}
\providecommand{\doi}[1]{\url{https://doi.org/#1}}
\bibcommenthead

\bibitem{ashcroft2011}
\bibinfo{author}{Ashcroft, N.} \& \bibinfo{author}{Mermin, N.}
\newblock \emph{\bibinfo{title}{Solid {S}tate {P}hysics}}
  (\bibinfo{publisher}{Cengage Learning}, \bibinfo{year}{2011}).

\bibitem{tsui1982}
\bibinfo{author}{Tsui, D.~C.}, \bibinfo{author}{Stormer, H.~L.} \&
  \bibinfo{author}{Gossard, A.~C.}
\newblock \bibinfo{title}{Two-{D}imensional {M}agnetotransport in the {E}xtreme
  {Q}uantum {L}imit}.
\newblock \emph{\bibinfo{journal}{Phys. Rev. Lett.}}
  \textbf{\bibinfo{volume}{48}}, \bibinfo{pages}{1559--1562}
  (\bibinfo{year}{1982}).
\newblock \doi{10.1103/PhysRevLett.48.1559} .

\bibitem{fenton1968}
\bibinfo{author}{Fenton, E.~W.}
\newblock \bibinfo{title}{Excitonic {I}nsulator in a {M}agnetic {F}ield}.
\newblock \emph{\bibinfo{journal}{Phys. Rev.}} \textbf{\bibinfo{volume}{170}},
  \bibinfo{pages}{816--821} (\bibinfo{year}{1968}).
\newblock \doi{10.1103/PhysRev.170.816} .

\bibitem{yoshioka1981}
\bibinfo{author}{Yoshioka, D.} \& \bibinfo{author}{Fukuyama, H.}
\newblock \bibinfo{title}{Electronic {P}hase {T}ransition of {G}raphite in a
  {S}trong {M}agnetic {F}ield}.
\newblock \emph{\bibinfo{journal}{J. Phys. Soc. Japan}}
  \textbf{\bibinfo{volume}{50}}~(3), \bibinfo{pages}{725--726}
  (\bibinfo{year}{1981}).
\newblock \doi{10.1143/JPSJ.50.725} .

\bibitem{halperin1987}
\bibinfo{author}{Halperin, B.~I.}
\newblock \bibinfo{title}{Possible {S}tates for a {T}hree-{D}imensional
  {E}lectron {G}as in a {S}trong {M}agnetic {F}ield}.
\newblock \emph{\bibinfo{journal}{Jpn. J. Appl. Phys.}}
  \textbf{\bibinfo{volume}{26}}~(S3-3), \bibinfo{pages}{1913}
  (\bibinfo{year}{1987}).
\newblock \doi{10.7567/JJAPS.26S3.1913} .

\bibitem{fradkin1999}
\bibinfo{author}{Fradkin, E.} \& \bibinfo{author}{Kivelson, S.~A.}
\newblock \bibinfo{title}{Liquid-crystal phases of quantum hall systems}.
\newblock \emph{\bibinfo{journal}{Phys. Rev. B}} \textbf{\bibinfo{volume}{59}},
  \bibinfo{pages}{8065--8072} (\bibinfo{year}{1999}).
\newblock \doi{10.1103/PhysRevB.59.8065} .

\bibitem{khveshchenko2001}
\bibinfo{author}{Khveshchenko, D.~V.}
\newblock \bibinfo{title}{Magnetic-{F}ield-{I}nduced {I}nsulating {B}ehavior in
  {H}ighly {O}riented {P}yrolitic {G}raphite}.
\newblock \emph{\bibinfo{journal}{Phys. Rev. Lett.}}
  \textbf{\bibinfo{volume}{87}}, \bibinfo{pages}{206401}
  (\bibinfo{year}{2001}).
\newblock \doi{10.1103/PhysRevLett.87.206401} .

\bibitem{jain2007}
\bibinfo{author}{Jain, J.~K.}
\newblock \emph{\bibinfo{title}{Composite {F}ermions}}
  (\bibinfo{publisher}{Cambridge University Press}, \bibinfo{year}{2007}).

\bibitem{edwards1996}
\bibinfo{author}{Edwards, D.~M.} \& \bibinfo{author}{Green, A. C.~M.}
\newblock \bibinfo{title}{Heavy fermions in high magnetic fields}.
\newblock \emph{\bibinfo{journal}{Z. Phys. B}}
  \textbf{\bibinfo{volume}{103}}~(2), \bibinfo{pages}{243--249}
  (\bibinfo{year}{1996}).
\newblock \doi{10.1007/s002570050367} .

\bibitem{harrison2011}
\bibinfo{author}{Altarawneh, M.~M.} \emph{et~al.}
\newblock \bibinfo{title}{Sequential {S}pin {P}olarization of the {F}ermi
  {S}urface {P}ockets in {U}{R}u$_{2}${S}i$_{2}$ and {I}ts {I}mplications for
  the {H}idden {O}rder}.
\newblock \emph{\bibinfo{journal}{Phys. Rev. Lett.}}
  \textbf{\bibinfo{volume}{106}}, \bibinfo{pages}{146403}
  (\bibinfo{year}{2011}).
\newblock \doi{10.1103/PhysRevLett.106.146403} .

\bibitem{Kushwaha2019}
\bibinfo{author}{Kushwaha, S.~K.} \emph{et~al.}
\newblock \bibinfo{title}{Magnetic field-tuned {F}ermi liquid in a {K}ondo
  insulator}.
\newblock \emph{\bibinfo{journal}{Nature Communications}}
  \textbf{\bibinfo{volume}{10}}~(1), \bibinfo{pages}{5487}
  (\bibinfo{year}{2019}).
\newblock \doi{10.1038/s41467-019-13421-w} .

\bibitem{iye1982}
\bibinfo{author}{Iye, Y.} \emph{et~al.}
\newblock \bibinfo{title}{High-magnetic-field electronic phase transition in
  graphite observed by magnetoresistance anomaly}.
\newblock \emph{\bibinfo{journal}{Phys. Rev. B}} \textbf{\bibinfo{volume}{25}},
  \bibinfo{pages}{5478--5485} (\bibinfo{year}{1982}).
\newblock \doi{10.1103/PhysRevB.25.5478} .

\bibitem{zhu2019}
\bibinfo{author}{Zhu, Z.} \emph{et~al.}
\newblock \bibinfo{title}{Graphite in 90 {T}: {E}vidence for
  {S}trong-{C}oupling {E}xcitonic {P}airing}.
\newblock \emph{\bibinfo{journal}{Phys. Rev. X}} \textbf{\bibinfo{volume}{9}},
  \bibinfo{pages}{011058} (\bibinfo{year}{2019}).
\newblock \doi{10.1103/PhysRevX.9.011058} .

\bibitem{ramshaw2018}
\bibinfo{author}{Ramshaw, B.~J.} \emph{et~al.}
\newblock \bibinfo{title}{Quantum limit transport and destruction of the {W}eyl
  nodes in {T}a{A}s}.
\newblock \emph{\bibinfo{journal}{Nat. Comm.}}
  \textbf{\bibinfo{volume}{9}}~(1), \bibinfo{pages}{2217}
  (\bibinfo{year}{2018}).
\newblock \doi{10.1038/s41467-018-04542-9} .

\bibitem{geim2007}
\bibinfo{author}{Geim, A.~K.} \& \bibinfo{author}{Novoselov, K.~S.}
\newblock \bibinfo{title}{The rise of graphene}.
\newblock \emph{\bibinfo{journal}{Nat. Mater.}}
  \textbf{\bibinfo{volume}{6}}~(3), \bibinfo{pages}{183--191}
  (\bibinfo{year}{2007}).
\newblock \doi{10.1038/nmat1849} .

\bibitem{goerbig2011}
\bibinfo{author}{Goerbig, M.~O.}
\newblock \bibinfo{title}{Electronic properties of graphene in a strong
  magnetic field}.
\newblock \emph{\bibinfo{journal}{Rev. Mod. Phys.}}
  \textbf{\bibinfo{volume}{83}}, \bibinfo{pages}{1193--1243}
  (\bibinfo{year}{2011}).
\newblock \doi{10.1103/RevModPhys.83.1193} .

\bibitem{moon2012}
\bibinfo{author}{Moon, P.} \& \bibinfo{author}{Koshino, M.}
\newblock \bibinfo{title}{Energy spectrum and quantum {H}all effect in twisted
  bilayer graphene}.
\newblock \emph{\bibinfo{journal}{Phys. Rev. B}} \textbf{\bibinfo{volume}{85}},
  \bibinfo{pages}{195458} (\bibinfo{year}{2012}).
\newblock \doi{10.1103/PhysRevB.85.195458} .

\bibitem{Terashima2017}
\bibinfo{author}{Terashima, T.~T.} \emph{et~al.}
\newblock \bibinfo{title}{Magnetization {P}rocess of the {K}ondo {I}nsulator
  {Y}b{B}$_{12}$ in {U}ltrahigh {M}agnetic {F}ields}.
\newblock \emph{\bibinfo{journal}{J. Phys. Soc. Japan}}
  \textbf{\bibinfo{volume}{86}}~(5), \bibinfo{pages}{054710}
  (\bibinfo{year}{2017}).
\newblock \doi{10.7566/JPSJ.86.054710} .

\bibitem{SI}
\bibinfo{title}{See supplementary information}  .

\bibitem{XiangScience}
\bibinfo{author}{Xiang, Z.} \emph{et~al.}
\newblock \bibinfo{title}{Quantum oscillations of electrical resistivity in an
  insulator}.
\newblock \emph{\bibinfo{journal}{Science}}
  \textbf{\bibinfo{volume}{362}}~(6410), \bibinfo{pages}{65--69}
  (\bibinfo{year}{2018}).
\newblock \doi{10.1126/science.aap9607} .

\bibitem{Xiang2021}
\bibinfo{author}{Xiang, Z.} \emph{et~al.}
\newblock \bibinfo{title}{Unusual high-field metal in a {K}ondo insulator}.
\newblock \emph{\bibinfo{journal}{Nat. Phys.}}
  \textbf{\bibinfo{volume}{17}}~(7), \bibinfo{pages}{788--793}
  (\bibinfo{year}{2021}).
\newblock \doi{10.1038/s41567-021-01216-0} .

\bibitem{liu2022}
\bibinfo{author}{Liu, H.} \emph{et~al.}
\newblock \bibinfo{title}{f-electron hybridised {F}ermi surface in magnetic
  field-induced metallic {Y}b{B}$_{12}$}.
\newblock \emph{\bibinfo{journal}{npj Quantum Mater.}}
  \textbf{\bibinfo{volume}{7}}~(1), \bibinfo{pages}{12} (\bibinfo{year}{2022}).
\newblock \doi{10.1038/s41535-021-00413-7} .

\bibitem{Riseborough2000}
\bibinfo{author}{Riseborough, P.~S.}
\newblock \bibinfo{title}{Heavy fermion semiconductors}.
\newblock \emph{\bibinfo{journal}{Adv. Phys.}}
  \textbf{\bibinfo{volume}{49}}~(3), \bibinfo{pages}{257--320}
  (\bibinfo{year}{2000}).
\newblock \doi{10.1080/000187300243345} .

\bibitem{Sugiyama1988}
\bibinfo{author}{Sugiyama, K.}, \bibinfo{author}{Iga, F.},
  \bibinfo{author}{Kasaya, M.}, \bibinfo{author}{Kasuya, T.} \&
  \bibinfo{author}{Date, M.}
\newblock \bibinfo{title}{Field-{I}nduced {M}etallic {S}tate in {Y}b{B}$_{12}$
  under {H}igh {M}agnetic {F}ield}.
\newblock \emph{\bibinfo{journal}{J. Phys. Soc. Japan}}
  \textbf{\bibinfo{volume}{57}}~(11), \bibinfo{pages}{3946--3953}
  (\bibinfo{year}{1988}).
\newblock \doi{10.1143/JPSJ.57.3946} .

\bibitem{gDefinition}
\bibinfo{title}{The convention used throughout this work is $g^* = g_\mathrm{J}
  m_\mathrm{J}$}  .

\bibitem{Moll2017}
\bibinfo{author}{Moll, P. J.~W.} \emph{et~al.}
\newblock \bibinfo{title}{Emergent magnetic anisotropy in the cubic
  heavy-fermion metal {C}e{I}n$_{3}$}.
\newblock \emph{\bibinfo{journal}{npj Quantum Materials}}
  \textbf{\bibinfo{volume}{2}}~(1), \bibinfo{pages}{46} (\bibinfo{year}{2017}).
\newblock \doi{10.1038/s41535-017-0052-5} .

\bibitem{Kurihara2021}
\bibinfo{author}{Kurihara, R.} \emph{et~al.}
\newblock \bibinfo{title}{Field-induced valence fluctuations in
  {Y}b{B}$_{12}$}.
\newblock \emph{\bibinfo{journal}{Phys. Rev. B}}
  \textbf{\bibinfo{volume}{103}}, \bibinfo{pages}{115103}
  (\bibinfo{year}{2021}).
\newblock \doi{10.1103/PhysRevB.103.115103} .

\bibitem{Iga1998}
\bibinfo{author}{Iga, F.}, \bibinfo{author}{Shimizu, N.} \&
  \bibinfo{author}{Takabatake, T.}
\newblock \bibinfo{title}{Single crystal growth and physical properties of
  kondo insulator {Y}b{B}$_{12}$}.
\newblock \emph{\bibinfo{journal}{J. Magn. Magn. Mater.}}
  \textbf{\bibinfo{volume}{177-181}}, \bibinfo{pages}{337--338}
  (\bibinfo{year}{1998}).
\newblock \doi{10.1016/S0304-8853(97)00493-9} .

\bibitem{Shoenberg1984}
\bibinfo{author}{Shoenberg, D.}
\newblock \emph{\bibinfo{title}{Magnetic {O}scillations in {M}etals}} Cambridge
  Monographs on Physics (\bibinfo{publisher}{Cambridge University Press},
  \bibinfo{year}{1984}).

\bibitem{Coffey2000}
\bibinfo{author}{Coffey, T.} \emph{et~al.}
\newblock \bibinfo{title}{Measuring radio frequency properties of materials in
  pulsed magnetic fields with a tunnel diode oscillator}.
\newblock \emph{\bibinfo{journal}{Rev. Sci. Instrum.}}
  \textbf{\bibinfo{volume}{71}}~(12), \bibinfo{pages}{4600--4606}
  (\bibinfo{year}{2000}).
\newblock \doi{10.1063/1.1321301} .

\bibitem{Xiang2022}
\bibinfo{author}{Xiang, Z.} \emph{et~al.}
\newblock \bibinfo{title}{Hall {A}nomaly, {Q}uantum {O}scillations and
  {P}ossible {L}ifshitz {T}ransitions in {K}ondo {I}nsulator {Y}b{B}$_{12}$:
  {E}vidence for {U}nconventional {C}harge {T}ransport}.
\newblock \emph{\bibinfo{journal}{Phys. Rev. X}} \textbf{\bibinfo{volume}{12}},
  \bibinfo{pages}{021050} (\bibinfo{year}{2022}).
\newblock \doi{10.1103/PhysRevX.12.021050} .

\bibitem{Aoki2013}
\bibinfo{author}{Aoki, D.}, \bibinfo{author}{Knafo, W.} \&
  \bibinfo{author}{Sheikin, I.}
\newblock \bibinfo{title}{Heavy fermions in a high magnetic field}.
\newblock \emph{\bibinfo{journal}{Comptes Rendus Physique}}
  \textbf{\bibinfo{volume}{14}}~(1), \bibinfo{pages}{53--77}
  (\bibinfo{year}{2013}).
\newblock \doi{10.1016/j.crhy.2012.11.004} .

\bibitem{okamura2005}
\bibinfo{author}{Okamura, H.} \emph{et~al.}
\newblock \bibinfo{title}{Indirect and {D}irect {E}nergy {G}aps in {K}ondo
  {S}emiconductor {Y}b{B}$_{12}$}.
\newblock \emph{\bibinfo{journal}{J. Phys. Soc. Japan}}
  \textbf{\bibinfo{volume}{74}}~(7), \bibinfo{pages}{1954--1957}
  (\bibinfo{year}{2005}).
\newblock \doi{10.1143/JPSJ.74.1954} .

\bibitem{Iga2010}
\bibinfo{author}{Iga, F.} \emph{et~al.}
\newblock \bibinfo{title}{Anisotropic magnetoresistance and collapse of the
  energy gap in {Y}b$_{1-x}${L}u$_{x}${B}$_{12}$}.
\newblock \emph{\bibinfo{journal}{J. Phys. Conf. Ser.}}
  \textbf{\bibinfo{volume}{200}}~(1), \bibinfo{pages}{012064}
  (\bibinfo{year}{2010}).
\newblock \doi{10.1088/1742-6596/200/1/012064} .

\bibitem{Knolle2015}
\bibinfo{author}{Knolle, J.} \& \bibinfo{author}{Cooper, N.~R.}
\newblock \bibinfo{title}{Quantum {O}scillations without a {F}ermi {S}urface
  and the {A}nomalous de {H}aas--van {A}lphen {E}ffect}.
\newblock \emph{\bibinfo{journal}{Phys. Rev. Lett.}}
  \textbf{\bibinfo{volume}{115}}, \bibinfo{pages}{146401}
  (\bibinfo{year}{2015}).
\newblock \doi{10.1103/PhysRevLett.115.146401} .

\bibitem{Zhang2016}
\bibinfo{author}{Zhang, L.}, \bibinfo{author}{Song, X.-Y.} \&
  \bibinfo{author}{Wang, F.}
\newblock \bibinfo{title}{Quantum oscillation in narrow-gap topological
  insulators}.
\newblock \emph{\bibinfo{journal}{Phys. Rev. Lett.}}
  \textbf{\bibinfo{volume}{116}}, \bibinfo{pages}{046404}
  (\bibinfo{year}{2016}).
\newblock \doi{10.1103/PhysRevLett.116.046404} .

\bibitem{Knolle2017}
\bibinfo{author}{Knolle, J.} \& \bibinfo{author}{Cooper, N.~R.}
\newblock \bibinfo{title}{Excitons in topological kondo insulators: Theory of
  thermodynamic and transport anomalies in {S}m{B}$_6$}.
\newblock \emph{\bibinfo{journal}{Phys. Rev. Lett.}}
  \textbf{\bibinfo{volume}{118}}, \bibinfo{pages}{096604}
  (\bibinfo{year}{2017}).
\newblock \doi{10.1103/PhysRevLett.118.096604} .

\bibitem{He2021}
\bibinfo{author}{He, W.-Y.} \& \bibinfo{author}{Lee, P.~A.}
\newblock \bibinfo{title}{Quantum oscillation of thermally activated
  conductivity in a monolayer {W}{T}e$_{2}$-like excitonic insulator}.
\newblock \emph{\bibinfo{journal}{Phys. Rev. B}}
  \textbf{\bibinfo{volume}{104}}, \bibinfo{pages}{L041110}
  (\bibinfo{year}{2021}).
\newblock \doi{10.1103/PhysRevB.104.L041110} .

\bibitem{Coleman1994}
\bibinfo{author}{Coleman, P.}, \bibinfo{author}{Miranda, E.} \&
  \bibinfo{author}{Tsvelik, A.}
\newblock \bibinfo{title}{Odd-frequency pairing in the {K}ondo lattice}.
\newblock \emph{\bibinfo{journal}{Phys. Rev. B}} \textbf{\bibinfo{volume}{49}},
  \bibinfo{pages}{8955--8982} (\bibinfo{year}{1994}).
\newblock \doi{10.1103/PhysRevB.49.8955} .

\bibitem{Baskaran2015}
\bibinfo{author}{Baskaran, G.}
\newblock \bibinfo{title}{{M}ajorana {F}ermi {S}ea in {I}nsulating {S}m{B}$_6$:
  {A} proposal and a {T}heory of {Q}uantum {O}scillations in {K}ondo
  {I}nsulators} (\bibinfo{year}{2015}).
\newblock \bibinfo{note}{Preprint at \url{https://arxiv.org/abs/1507.03477}}.

\bibitem{Chowdhury2018}
\bibinfo{author}{Chowdhury, D.}, \bibinfo{author}{Sodemann, I.} \&
  \bibinfo{author}{Senthil, T.}
\newblock \bibinfo{title}{Mixed-valence insulators with neutral {F}ermi
  surfaces}.
\newblock \emph{\bibinfo{journal}{Nat. Comm.}}
  \textbf{\bibinfo{volume}{9}}~(1), \bibinfo{pages}{1766}
  (\bibinfo{year}{2018}).
\newblock \doi{10.1038/s41467-018-04163-2} .

\bibitem{Sodemann2018}
\bibinfo{author}{Sodemann, I.}, \bibinfo{author}{Chowdhury, D.} \&
  \bibinfo{author}{Senthil, T.}
\newblock \bibinfo{title}{Quantum oscillations in insulators with neutral
  {F}ermi surfaces}.
\newblock \emph{\bibinfo{journal}{Phys. Rev. B}} \textbf{\bibinfo{volume}{97}},
  \bibinfo{pages}{045152} (\bibinfo{year}{2018}).
\newblock \doi{10.1103/PhysRevB.97.045152} .

\bibitem{Shen2018}
\bibinfo{author}{Shen, H.} \& \bibinfo{author}{Fu, L.}
\newblock \bibinfo{title}{Quantum {O}scillation from {I}n-{G}ap {S}tates and a
  {N}on-{H}ermitian {L}andau {L}evel {P}roblem}.
\newblock \emph{\bibinfo{journal}{Phys. Rev. Lett.}}
  \textbf{\bibinfo{volume}{121}}, \bibinfo{pages}{026403}
  (\bibinfo{year}{2018}).
\newblock \doi{10.1103/PhysRevLett.121.026403} .

\end{thebibliography}



\end{document}


\title[Supplementary Information]{Supplementary Information}

\maketitle

\section{Methods}
\subsection{Growth and Structural Characterization}
Single crystals of YbB$_{12}$ were grown via the traveling solvent (TS) method under Ar gas atmosphere in 300 bar high pressure floating zone furnace (SciDre GmbH) ~\cite{Phelan2018} at the Platform for the Accelerated Realization, Analysis and Discovery of Interface Materials (PARADIM) user facility at Johns Hopkins University using a similar procedure as previously documented~\cite{Iga1998}. Subsequent crystal structure analysis employed using Laue and single crystal X-ray diffraction. Single crystal domains used for the high field measurements were cut from a larger boule piece upon being oriented using a Photonic Science Laue diffractometer (Fig. ~\ref{SI_Laue}).  A smaller single crystal of YbB$_{12}$ than those used for Laue diffraction was mounted onto a loop fiber of a goniometer using n-grease, and the goniometer was placed on a Bruker D8 Venture X-ray diffractometer equipped with Ag K$\alpha$ ($\lambda$ = 0.56086 $\rm{\AA}$) radiation. The crystallographic lattice parameters, relevant data collection information, and the corresponding refinement statistics are found in Table~\ref{SI_tab_XRD}. The cubic Laue symmetry of m-3m and the observed systematic absences led to a space group selection of Fm-3m. Structure refinement was conducted using SHELX2018~\cite{Sheldrick2015,Krause2015}, where in corresponding order, the atom positions were refined, anisotropic motion was taken into account, an extinction correction was applied, and a weighting scheme was allowed to converge. Finally, it is important to note that all data were corrected for absorption using the multi-scan method. These diffraction methods and subsequent analysis show that the TS grown crystal was of a sufficient quality needed for this study.

\begin{figure*}[!h]%
\centering
\includegraphics[width=0.75\textwidth]{SI_Laue.png}
\caption{A single crystal domain on edge oriented close to the [011] crystallographic axis.}\label{SI_Laue}
\end{figure*}

\begin{table}
\caption{Crystallographic and refinement parameters for YbB$_{12}$.}\label{SI_tab_XRD}
\begin{center}
\begin{tabular}{ c c c }
\toprule
 a ($\rm{\AA}$) & 7.4752(2) \\ 
 V ($\rm{\AA}^{3}$) & 417.70(3) \\  
 Z & 4  \\
 T (K) & 297(2)  \\
 $\theta$-range ($\deg$) & 3.7 - 68.1 \\
 $\mu$ (mm$^{-1}$) & 11.93 \\
 Measured reflections & 14740 \\
 Independent reflections & 273 \\
 $R_{int}$ & 0.072 \\
 $\Delta\rho_{\rm{max}}$ (e/$\rm{\AA}^3$) & 1.23 \\
 $\Delta\rho_{\rm{min}}$ (e/$\rm{\AA}^3$) & 1.51 \\
 Extinction coefficient & 0.0088(5) \\
 Data / restraints / parameters & 273 / 0 / 7 \\
 R$_{\rm{1}}$ ($F^2 > 2\sigma(F^2)$) & 0.010 \footnotemark[1] \\
 wR$_{\rm{2}}$ ($F^2$) & 0.020 \footnotemark[2] \\
 \bottomrule
\end{tabular}
\end{center}
\footnotetext[a]{ $R_1 = \frac{\sum \big\vert \vert F_o \vert - \vert F_c \vert \big\vert}{\sum \vert F_o \vert}$}

\footnotetext[b]{ $wR_2 = \left( \frac{\sum w(F_o^2 - F_c^2)^2 }{\sum w(F_o^2)^2} \right)^{1/2}$}

\end{table}

\subsection{Pulsed Field Experiments}
Electrical transport and tunnel-diode oscillator (TDO) measurements were performed in the National High Magnetic Field Laboratory Pulsed Field Facility at Los Alamos National Laboratory. Experiments utilized the 65T and 75T duplex magnets. The sample was immersed in $^4$He liquid for experiments between 1.5K and 4K, and $^3$He liquid for experiments below 1.5K. Temperatures above 4K utilized $^4$He gas.

Electrical transport was performed using a conventional four-wire, lock-in technique ($\sim$265kHz). Contacts were made using Au pads and reinforced with silver paste. TDO~\cite{Coffey2000} experiments were performed on the same sample as was used for conventional electrical transport. For this experiment, a 13 turn coil (46 gauge high-purity copper wire) was tightly wrapped around the sample. GE varnish was used to secure the coil. The TDO circuit connecting to the coil around the sample formed a tank circuit with a resonant frequency of $\sim$30MHz at measurement conditions. Mixing was used to bring the measured frequency down to $\sim$2MHz.

\section{Physical Property Characterization}
The zero-field resistivity of the YbB$_{12}$ sample used for the high field studies reported in the main text was first characterized using a low-frequency AC resistance bridge and a standard four-wire configuration in a Quantum Design Physical Property Measurement System (PPMS). The resistivity increased by over 4 orders of magnitude when cooled from 300K to 0.5K, as shown in Fig. \ref{SI_resistivity}. The insulating behavior stems from the opening of small gaps ($E_a = 2.7$ meV and $E_a = 4.7$ meV, inset). These gaps are thought to arise from the hybridization between conduction electrons and largely localized $f$ -electrons~\cite{Riseborough2000}. Below $\sim$3K the resistivity exhibits a ``plateau" seen in other YbB$_{12}$ samples, often attributed to a metallic surface~\cite{Iga1998,XiangScience,Sato2021}.

\begin{figure*}[!h]%
\centering
\includegraphics[width=0.5\textwidth]{SI_resistivity.eps}
\caption{Resistivity as a function of temperature in zero magnetic field. The resistivity increases by a factor of 2.8$\times10^{4}$ from 300K to 0.5K owing to the opening of gaps (inset).}\label{SI_resistivity}
\end{figure*}

Figure \ref{SI_susceptibility} shows the magnetic susceptibility as a function of temperature from 1.8K to 350K measured with magnetic fields of 0.1T and 5T. The data were obtained on a Quantum Design Magnetic Properties Measurement System (MPMS3) equipped with a 7T magnet. Above $\sim$170K the susceptibility exhibits Curie-Weiss behavior (Figure \ref{SI_susceptibility}b) giving a Curie temperature of -134.8$\pm$1.3K (-123.6$\pm$2.3K) and effective moments per Yb of 4.6$\pm$0.3$\mu_B$ (4.4$\pm$0.3$\mu_B$) for an applied field of 0.1T (5T). These values are consistent with a nominal +3 valence for Yb (effective moment of 4.54 $\mu_B$)~\cite{Iga1998}. The susceptibility exhibits a maximum at $\sim$75K, minimum at $\sim$17K, and a low temperature increase which is suppressed by field. The maximum is characteristic of Kondo singlet formation and the temperature of the susceptibility maximum is proportional to the Kondo temperature~\cite{Aoki2013}. The low temperature increase in susceptibility is indicative of magnetic impurities~\cite{Iga1998}.

\begin{figure*}[!h]%
\centering
\includegraphics[width=0.9\textwidth]{SI_susceptibility.eps}
\caption{(a) Magnetic susceptibility as a function of temperature measured with fields of 0.1T (red) and 5T (blue). (b) Inverse susceptibility with Curie-Weiss fits.}\label{SI_susceptibility}
\end{figure*}

Specific heat measurements (Fig. \ref{SI_heat_capacity}) were collected in a PPMS calorimeter using a quasiadiabatic thermal relaxation technique with a $^{3}$He insert before and after the sample was cut and polished for the pulsed field experiments. Before polishing the mass of the crystal was $\sim$5 mg and after polishing the mass was $\sim$1.1 mg. Both measurements show a low temperature upturn, often attributed to a Schottky contribution~\cite{Sato2021}. In both cases a linear extrapolation of $C/T$ yields a Sommerfeld coefficient of 1.3$\pm$0.2 mJ/molK$^{2}$. This value is smaller than values reported in the literature~\cite{Sato2021}, but of the same order of magnitude.

\begin{figure*}[!h]%
\centering
\includegraphics[width=0.5\textwidth]{SI_heat_capacity.eps}
\caption{Heat capacity as a function of temperature before and after polishing. A linear extrapolation of $C/T$ gives an intercept of $\gamma = 1.3\pm0.2$mJ/molK$^{2}$.}\label{SI_heat_capacity}
\end{figure*}

\section{Reverse Quantum Limit Criterion}

Beginning from Eq. (2) in the main text, we have

\begin{equation}
E_C^\uparrow - \mu = \Delta_0 - \frac{1}{2} g^{*} \mu_{B} B + \frac{\hbar e}{m^{*}} B \left(\nu+\frac{1}{2}\right). \label{eq_rql}
\end{equation}

\noindent Landau level crossings occur when $E_C^\uparrow = \mu$. Therefore,

\begin{equation}
\Delta_0 = \left( \frac{1}{2} g^{*} \mu_{B} - \frac{\hbar e}{m^{*}} \left(\nu+\frac{1}{2}\right) \right) B.\label{eq_rql_gap}
\end{equation}

\noindent
Since the zero-field gap and field are positive quantities, the above equality requires 

\begin{equation}
\frac{1}{2} g^{*} \mu_{B} > \frac{\hbar e}{m^{*}} \left(\nu+\frac{1}{2}\right).
\end{equation}

\noindent Focusing on the first Landau level ($\nu = 0$) and substituting $\mu_B = \frac{e \hbar}{2 m_e}$ yields the criterion for the reverse quantum limit:

\begin{equation}
g^{*} \frac{m^{*}}{m_{e}} > 2.
\end{equation}

\noindent Landau level diagrams for the reverse quantum limit and the other four high-field scenarios described in Fig. 4 of the main text are provided in Fig.~\ref{SI_4cases.eps}.

\begin{figure*}[!h]%
\centering
\includegraphics[trim=50 0 75 0, clip, width=1\textwidth]{SI_4cases.eps}
\caption{Four high-field scenarios described in Fig. 4 of the main text corresponding to whether the material is gapped at the chemical potential and the relative size of the Zeeman and cyclotron energies. Colors correspond to the different Landau indices defined in the first panel and dashed/solid lines denote different spins.}\label{SI_4cases.eps}
\end{figure*}

\section{Landau Level Pinning to Insulator-Metal Transition in Reverse Quantum Limit}

Landau level crossings occur when $E_C^\uparrow - \mu = 0$. If we assume $g^*$ and $m^*$ are independent of field we can write the following expression for Landau level crossings

\begin{equation}
    0 = \Delta(B) - \frac{1}{2} g^* \mu_B B + \frac{\hbar e}{m^*} B (\nu + \frac{1}{2}) \label{eq_LL_crossing}
\end{equation}

\noindent We can simplify this expression by relating $g^*$ to the insulator-metal transition field. To do this we consider that at the insulator-metal transition, the activation energy is zero, so

\begin{equation}
    2E_a = (E_C^\uparrow - \mu) - (E_V^\downarrow - \mu) = 0.
\end{equation}

\noindent Assuming $\Delta(B) = \Delta_0$ is a constant in the insulating state prior to the insulator-metal transition, and recognizing that the activation energy in the reverse quantum limit is set by excitations across the $\nu=0$ Landau levels, we can write

\begin{equation}
    2\Delta_0 - g^*\mu_B B_{IM} + \frac{\hbar e}{m^*} B_{IM} = 0 
\end{equation}

\noindent This expression relates the zero-field gap to the insulator-metal transition field and effective $g$ factor. It shows Landau quantization will tend to push the insulator-metal transition to a higher field than if only the Zeeman energy is considered.

\begin{equation}
    B_{IM} = \frac{2\Delta_0}{g^*\mu_B  - \frac{\hbar e}{m^*}}
\end{equation}

\noindent If we instead solve the above equation for $g^*\mu_B$ we have 

\begin{equation}
     g^*\mu_B = \frac{2\Delta_0}{ B_{IM} } + \frac{\hbar e}{m^*} .
\end{equation}

\noindent Substituting this into Eq.(\ref{eq_LL_crossing}) gives

\begin{equation}
    0 = \Delta(B) - \frac{1}{2} \left( \frac{2\Delta_0}{ B_{IM} } + \frac{\hbar e}{m^*}\right) B + \frac{\hbar e}{m^*} B (\nu + \frac{1}{2})
\end{equation}

\noindent which reduces 

\begin{equation}
    \nu \frac{\hbar e}{m^*} = \frac{\Delta_0}{ B_{IM}} - \frac{\Delta(B)}{B}. \label{eq_nu_pinning}
\end{equation}

\noindent When $\Delta(B)$ does not depend on field (i.e, $\Delta(B)=\Delta_0$ which applies in the insulating state) we have 

\begin{equation}
    \nu \frac{\hbar e}{m^*\Delta_0} =  \frac{1}{B_{IM}} - \frac{1}{B_{\nu}} \label{eq_nu_pinning_insulator}
\end{equation}

\noindent where we now explicitly add the subscript $\nu$ to $B$ to denote the field at which Landua level $\nu$ crosses the chemical potential. Since the right hand side of the equation is only weakly dependent on angle, in the reverse quantum limit quantum oscillations are expected to be pinned to the insulator-metal transition as is shown in Fig. 3 of the main text. We can also confirm this behavior by looking at the angular dependence of the quantum oscillation frequency in the insulating state because Eq.(\ref{eq_nu_pinning_insulator}) implies a constant quantum oscillation frequency which is independent of angle (Fig.~\ref{SI_fig_frequency_index}).

When the zero-field gap changes with field (relevant in the metallic state), Eq.(\ref{eq_nu_pinning}) applies. In this situation we expect the Landau levels to be pinned to the insulator-metal transition close to the transition, with deviations at higher field owing to $\Delta_0$ having a field dependence. This explains the slight deviations from the angular dependence of the insulator-metal transition in the metallic state shown in Fig. 3 of the main text. These deviations are more clearly shown by plotting the Landau index as a function of inverse field at fixed angles (Fig.~\ref{SI_fig_frequency_index}). As we show in subsequent sections, this behavior is consistent with the field-dependent quantum oscillation frequency in the metallic state and are related to anisotropies in the magnetization and Fermi surface.

\begin{figure*}[!h]%
\centering
\includegraphics[trim=50 0 75 0, clip, width=1\textwidth]{SI_fig_frequency_index.eps}
\caption{(a) Quantum oscillation frequency in the insulating state as a function of angle has minimal variations over a wide angular range. (b) Landau index as a function of inverse field in the metallic state (both quantities are referenced to the first observed quantum oscillation according to Eq.(\ref{eq_nu_pinning})). The angular dependence is related to anisotropies in the magnetization and Fermi surface.}\label{SI_fig_frequency_index}
\end{figure*}

\section{Field-Dependent Frequency in Reverse Quantum Limit}
Beginning from a generalization of Eq. (\ref{eq_rql}) where we do not assume a particular field dependence for the gap

\begin{equation}
    E_C^\uparrow - \mu = \Delta(B) -\frac{1}{2} g^*\mu_B B + \frac{\hbar e}{m^{*}} B \left(\nu+\frac{1}{2}\right),
\end{equation}

\noindent we impose the condition for quantum oscillations and solve for the Landau index

\begin{equation}
     \nu+\frac{1}{2} = \frac{g^* \mu_B m^{*}}{2 \hbar e} -\frac{\Delta(B) m^{*}}{\hbar e B}.
\end{equation}

Given that the quantum oscillation frequency is related to the Landau index~\cite{Shoenberg1984} via

\begin{equation}
    F = \frac{d\nu}{d\frac{1}{B}} = -B^2 \frac{d\nu}{dB} \label{eq_freq_QO},
\end{equation}

\noindent we find that in the reverse quantum limit

\begin{equation}
    F = \frac{m^*}{\hbar e} \left(B \frac{d\Delta(B)}{dB}  - \Delta(B)\right).
\end{equation}

More generally, starting from 

\begin{equation}
    E_C^\uparrow - \mu = \Delta(B) -\frac{1}{2} g^{*}\mu_B B + \frac{\hbar e}{m^{*}} B \left(\nu+\frac{1}{2}\right),
\end{equation}

\noindent but we let $g^{*}$, $m^{*}$, and $\Delta$ all vary with $B$. Then, we once again impose the condition for quantum oscillations and solve for the Landau index yielding

\begin{equation}
     \nu+\frac{1}{2} = \frac{m^{*}(B) g^{*}(B) \mu_B}{2\hbar e} -\frac{\Delta(B) m^{*}(B)}{\hbar e B}.
\end{equation}

\noindent Using the relation between the quantum oscillation frequency and Landau indices given by Eq. (\ref{eq_freq_QO}), we find

\begin{equation}
\begin{split}
    F = \frac{1}{\hbar e} \biggl( - m^{*}(B) \Delta(B)  + B \left[m^{*}(B) \frac{d\Delta(B)}{dB} + \Delta(B) \frac{dm^{*}(B)}{dB} \right] - \\ \frac{\mu_B}{2} B^2 \left[m^{*}(B) \frac{dg^{*}(B)}{dB} + g^{*}(B)\frac{dm^{*}(B)}{dB} \right] \biggr)
\end{split}
\end{equation}

\noindent Since $\frac{\mu_B}{2} B^2 << B$ for the field ranges in these experiments, we can neglect the last term in this expression giving 

\begin{equation}
    F = \frac{1}{\hbar e} \biggl( - m^{*}(B) \Delta(B)  + B \left[m^{*}(B) \frac{d\Delta(B)}{dB} + \Delta(B) \frac{dm^{*}(B)}{dB} \right]  \biggr).
\end{equation}

\noindent Then, because $m^{*}$ experimentally scales linearly with field, a non-linear field dependence in the quantum oscillation frequency requires $\Delta$ to vary with field.

\section{Relating Fermi Surface Area, Quantum Oscillation Frequency, and Landau Indices to Magnetization}

As described in the main text, interpreting our experiments in terms of the reverse quantum limit model suggests the field-dependent quantum oscillation frequency in the metallic state occurs because the gap evolves as a non-linear function of field in the metallic state. A possible origin for a non-linear field dependence of the gap is that the $c\!-\!f$ hybridization responsible for the gap is altered by the magnetic field~\cite{Aoki2013}. This possibility is consistent with partial $f$-electron polarization leading to non-linear magnetization in the field-induced metallic state~\cite{Sugiyama1988, Terashima2017, Xiang2021}. The non-linear magnetization used throughout our analysis is shown in Fig. \ref{SI_mag}.

\begin{figure*}[!h]%
\centering
\includegraphics[width=1\textwidth]{SI_mag.eps}
\caption{(a) Magnetization of YbB$_{12}$ with $\bf{H}\parallel$ [100] at T = 0.64K (blue). A linear fit to the magnetization in the insulating state (green) was used to obtain the non-linear magnetization. (b) The non-linear magnetization obtained by subtracting the linear fit from the total magnetization. Data taken from Ref.~\cite{Xiang2021}.}\label{SI_mag}
\end{figure*}

If this interpretation is correct, one expects the Fermi surface area measured by the quantum oscillations to be proportional to the non-linear portion of the magnetization corresponding to the partial $f$-electron polarization~\cite{Aoki2013}. More specifically, since magnetization is the magnetic dipole moment per unit volume one expects the Fermi surface area to be proportional to the non-linear magnetization to the $\frac{2}{3}$ power leading to

\begin{equation}
    A = a \left(\frac{M(B)}{\mu_B} \right)^{2/3} \label{eq_A_M}
\end{equation}

\noindent where $A$ is the Fermi surface area, $\mu_B$ is the Bohr magneton, $M(B)$ is the field-dependent, non-linear magnetization, and $a$ is a dimensionless constant related to the degeneracy factor (see next section). Similar analyses have been performed in heavy fermion systems including UPt$_3$~\cite{McCollam2021}.

Since the Fermi surface area is related to the quantum oscillation frequency ($F$) via the Onsager relation~\cite{Shoenberg1984},

\begin{equation}
    F = \frac{\hbar}{2 \pi e} A, 
\end{equation}

\noindent and $F$ is related to the Landau indices ($\nu$) through

\begin{equation}
    \nu = \frac{F}{B},
\end{equation}

\noindent it is possible to directly relate the Landau indices to the non-linear component of the magnetization. This relationship is

\begin{equation}
    \nu  = \frac{\hbar}{2 \pi e} \frac{1}{B} a \left(\frac{M(B)}{\mu_B} \right)^{2/3}\label{eq_LL_mag}.
\end{equation}

\noindent Using the relationship between the measured quantum oscillation frequency and the Landau index,

\begin{equation}
    F_{measured} = -B^2 \frac{d\nu}{dB},
\end{equation}

\noindent one can express the measured quantum oscillation frequency in terms of the non-linear magnetization.

\begin{equation}
    F_{measured} = -B^2 \frac{d}{dB}\left[\frac{\hbar}{2 \pi e} \frac{1}{B} a \left(\frac{M(B)}{\mu_B} \right)^{2/3}\right].\label{eq_freq}
\end{equation}

\noindent This is the expression used to fit the value of $a$ in the main text. Note, Eq. (\ref{eq_freq}) accounts for the effects of field-dependent quantum oscillations~\cite{Ruitenbeek1982, McCollam2013}.

Next, we relate the gap to the non-linear magnetization. From Eq.(\ref{eq_nu_pinning})

\begin{equation}
    \Delta(B) = \Delta_0 \frac{B}{B_{IM}} - \nu \frac{\hbar e}{m^*} B.
\end{equation}

\noindent Combining this condition with the relationship between the Landau indices and non-linear magnetization yields an the gap after the onset of non-linear magnetization, i.e. the insulator-metal transition:

\begin{equation}
    \Delta(B) = \Delta_0 \frac{B}{B_{IM}} - \frac{\hbar^2}{2 \pi m^*}a \left(\frac{M(B)}{\mu_B} \right)^{2/3}.
\end{equation}

\section{Estimating Carrier Density and Pocket Degeneracy}

Since the Fermi surface area is known from Eq. (\ref{eq_A_M}) it is possible to estimate the carrier concentration. Without knowing the exact shape of the Fermi surface, we first make a crude estimate for the carrier concentration assuming a spherical Fermi surface~\cite{ashcroft2011,Shoenberg1984}. In this case, the carrier concentration $n$ is related to the Fermi surface area through

\begin{equation} 
    n = \frac{A^{3/2}}{3\pi^2}.
\end{equation}

\noindent From Fig. 5b of the main text, the Fermi surface area near 60T is between $5-10$nm$^{-1}$, corresponding to a carrier concentration of $n\sim10^{20}-10^{21}$cm$^{-3}$ or $\sim0.2-0.5$ carriers per unit cell. This value is consistent with carrier concentrations of other Kondo metals~\cite{Kushwaha2019} and high field measurements we have performed on other YbB$_{12}$ samples at similar temperatures (Fig. \ref{SI_high_field_carrier}).

\begin{figure*}[!h]%
\centering
\includegraphics[width=0.5\textwidth]{SI_high_field_carrier.eps}
\caption{High field carrier concentration at T = 0.615K obtained on a different sample of YbB$_{12}$.}\label{SI_high_field_carrier}
\end{figure*}

Lastly, we express the constant $a$ introduced in Eq. (\ref{eq_A_M}) to the pocket degeneracy factor $D$, and in doing so provide an estimate for the pocket degeneracy. Our approach is to relate the magnetization contribution of the partially spin-polarized hybridized bands to the magnetization implied by the Landau level index in Eq. (\ref{eq_LL_mag}).

Similar to Pauli paramagnetism~\cite{ashcroft2011}, the magnetization per unit volume stemming from spin-polarization is given by
\begin{equation}
    M = \frac{1}{2} g^* \mu_B D n
\end{equation}

\noindent where $D$ is the pocket degeneracy and $n$ is the concentration of spin-polarized carriers. Accounting for spin degeneracy, $n$ is related to the wavevector $k$  through
\begin{equation}
    n = \frac{k^3}{3\pi^2}
\end{equation}

\noindent and $k$ is related to the Landau index via
\begin{equation}
    k^2 = \frac{2 e}{\hbar} B \left(\nu + \frac{1}{2} \right).
\end{equation}

\noindent Therefore, the magnetization is related to the Landau index through

\begin{equation}
M = \frac{1}{6\pi^2} g^* \mu_B \left( \frac{2 e}{\hbar} B \left(\nu + \frac{1}{2} \right) \right)^{3/2}\label{eq_LL_mag2}.
\end{equation}

\noindent Substituting Eq.(\ref{eq_LL_mag2}) into Eq. (\ref{eq_LL_mag}) and rearranging yields

\begin{equation}
    D = \frac{6 \pi^{7/2}}{g^* a^{3/2}}.
\end{equation}

\noindent Using the value of the fit parameter $a\approx3$ obtained from relating the measured frequency to the magnetization (i.e., Eq. (\ref{eq_freq})) and $g^*\approx2$, we find $D\approx 32$. This indicates YbB$_{12}$ has a Fermi surface with high degeneracy, similar to what has been suggested in Ref.~\cite{Liu2018,liu2022}. It is also important to note that even without exact knowledge of the pocket shape, there can be a number of additional factors affecting this degeneracy factor estimate such as the saturated moment per Yb differing from $\mu_{B}$ ~\cite{Terashima2017}, a change in the band Van Vleck contribution~\cite{Riseborough2000}, and/or a field-dependence of $g^*$.

\section{Determination of the Insulator-Metal Transition and Landau Level Indexing}
Values for the insulator-metal transition field as a function of angle were taken from the contactless resistivity measurements because the onset of the metallic state corresponds to a large change in the slope of the TDO frequency (e.g., Fig. 2 of main text). A precise determination of the insulator-metal transition was more difficult with the conventional resistivity measurements because of the dramatic change in the sample's resisitivity as a function of magnetic field: a current that would provide a sufficient signal-to-noise ratio to resolve the IM transition would cause large Joule heating in the insulating state. The use of the insulator-metal transition field from the contactless resistance measurements, and the slight differences in sample alignment between the contactless resistance and conventional resisitivity measurements, is responsible for the small difference in the concavity of the the angular dependences of the insulator-metal transition and insulating state quantum oscillations shown in Fig. 3 of the main text.

The angular dependence of the quantum oscillations was determined by assigning Landau level indices and following the angular dependence of each Landau level. We found this to be more robust than Fourier analysis owing to the few oscillations in the insulating state. Indexing in this manner makes it apparent that there is some evolution of fine structure in the insulating and metallic oscillations as a function of angle. In cases where peak splitting was apparent, we would take the average position of the split peaks as the field assigned to that Landau index. The origin of the peak splitting is unclear. While it has been suggested to be a consequence of spin-splitting~\cite{Xiang2021}, there will also be contributions from the shape of the Fermi surface, $g$-factor anisotropy, and anisotropic $f$-electron polarization~\cite{Terashima2017}, which complicate the situation. The exact origin of the peak splitting does not impact the central claims of this work.

\section{Lifshitz-Kosevich Analysis}
According to Lifshitz-Kosevich theory~\cite{Shoenberg1984}, quantum oscillations possess a temperature dependent amplitude damping factor ($R_T$) of

\begin{figure*}[!b]%
\centering
\includegraphics[width=0.9\textwidth]{SI_mass.png}
\caption{Effective mass analysis near [100] in the (a,c) insulating and (b,d) metallic states based on Lifshitz-Kosevich theory. The temperature dependent amplitudes of the (a) insulating state oscillations were obtained by background normalization and the temperature dependent amplitudes of the (d) metallic state oscillations were obtained by background subtraction. The temperature dependent amplitudes of the (c) insulating and (d) metallic quantum oscillation are well described the Lifshitz-Kosevich formula. Comparisons are made to LK analyses from Ref.~\cite{Xiang2021, Xiang2022}}\label{SI_mass}
\end{figure*}

\begin{equation}
    R_T = \frac{\alpha \frac{T}{B} \frac{m^*}{m_e}}{\sinh\left(\alpha \frac{T}{B} \frac{m^*}{m_e}\right)}\label{eq_LK},
\end{equation}

\noindent where 
\begin{equation}
\alpha = \frac{2\pi^2 k_B m_e}{e\hbar}.
\end{equation}

In order to isolate the amplitude of the quantum oscillations to determine $m^*$ from Eq.(\ref{eq_LK}), it is necessary to account for the background signal. A simple polynomial subtraction proved to be sufficient for the TDO data, however the significant magnetoresistance in the insulating state complicates the analysis. Following Pippard~\cite{Pippard1965,Shoenberg1984}, we perform a background normalization~\cite{Xiang2022} instead of a subtraction for the quantum oscillations in the insulating state, effectively assuming

\begin{equation}
    \frac{\rho_{oscillatory}}{\rho_{background}} \sim R_T \frac{g_{oscillatory}(\mu)}{g_{background}(\mu)}
\end{equation}

\noindent where $\rho$ is resistivity, $g(\mu)$ is the density of states at the chemical potential. This normalization more properly accounts for temperature dependent scattering rates.

The results of this analysis are shown in Fig. \ref{SI_mass}, along with data from LK analyses available in the literature~\cite{Xiang2021, Xiang2022}. The effective masses of both the insulating and metallic states are in good agreement with those previously reported~\cite{XiangScience, Xiang2021, Xiang2022}. We also performed a measurement of the mass as a function of field in the metallic state at 18.6$^{\circ}$ from the [100] crystallographic axis in the [100]-[011] plane (Fig. \ref{SI_mass_angles}). When combined with the measurements along the [100] axis, these measurements reveal minimal anisotropy in the effective mass of the field-induced metallic state.

\begin{figure*}[!h]%
\centering
\includegraphics[width=0.5\textwidth]{SI_mass_angles.eps}
\caption{Effective mass as a function of field at two different angles in the field-induced metallic state. At both angles the effective mass increases linearly with applied field. The effective mass exhibits minimal anisotropy. A comparison is made to data from Ref.~\cite{Xiang2021}}\label{SI_mass_angles}
\end{figure*}

While we did not observe strong deviations from LK behavior, this could be because we could not reach sufficiently low temperatures. Moreover, the substantial background magnetoresistance in the insulating state makes subtle deviations from LK difficult to discern. It is worth noting that the data presented in Ref.~\cite{Xiang2022} does show deviations from LK behavior in the SdH data. Although an alternative explanation is provided, it is possible the deviations could be caused by an excitonic insulator phase.

\section{Gap Extraction with Parallel Conduction Model}
To corroborate the gap closure results from the Arrhenius fits presented in Fig. 1 of the main text, we also analyzed the gap closure using a parallel conduction model that has previously been used to analyze transport in YbB$_{12}$~\cite{Sato2021}. The simplest model assumes the total conductance consists of two conduction channels, a constant channel often attributed to a metallic surface~\cite{Sato2021}, and an activated channel. As such, we fit the conductance as a function of temperature at fixed fields to the model

\begin{equation}
    G = A + G_0\exp\left(-\frac{E_a}{T}\right)
\end{equation}

\noindent where $G$ is the total conductance, $A$ is the constant conduction channel, $G_0$ is the activated channel's prefactor, and $E_a$ is the activation energy. Figure \ref{SI_compare_models} shows examples of fits to this model at fixed fields of 20T and 43T (applied parallel to [100]). Note, only the $E_a = 30K$ gap was included in this analysis because the larger gap (e.g., Fig. \ref{SI_resistivity}) has a negligible contribution to the total conductance in this temperature and field range.

In Fig. \ref{SI_compare_models}a, two fits are shown to the experimental data. The blue fit treats $E_a$ as a fit parameter. The red fit constrains $E_a$ to a fixed value obtained by assuming the gap evolves linearly in field from its measured zero field value (Fig. \ref{SI_resistivity}) through Zeeman splitting, i.e.

\begin{equation}
    E_a(B) = \Delta_0 - \frac{1}{2} g^* \mu_B B. \label{eq_zeeman}
\end{equation}

\begin{figure*}[!h]%
\centering
\includegraphics[trim=100 0 0 0, clip, width=1.1\textwidth]{SI_compare_models.eps}
\caption{Parallel conduction model fits to the conductance at fixed fields of (a) 20T and (b) 43T. In the lower field regime, the data is equally well fit with the gap considered to be a free parameter or constrained to take on a value assuming linear-in-field gap closure. The high field data cannot be described by linear-in-field gap closure.}\label{SI_compare_models}
\end{figure*}

While there is minimal difference between these two fits at 20T (Fig. \ref{SI_compare_models}a), the assumption of linear-in-field gap closure is unable to describe the data at high fields (Fig. \ref{SI_compare_models}b). Even with the addition of a third conduction channel which has been used to describe a possible nodal semimetallic state in SmB$_6$~\cite{Harrison2018}, such that

\begin{equation}
    G = A + G_0\exp\left(-\frac{E_a}{T}\right) + \beta T, \label{eq_conduction}
\end{equation}

\noindent it is clear the high field data cannot be fit by models which assume the gap closes linearly with field. Moreover, the gap values extracted from treating the gap as a free parameter are substantially (4-5$\times$) larger than those required were Eq. (\ref{eq_zeeman}) to apply over the entire field range.

Note, a comparison of the fits in Fig. \ref{SI_compare_models}b with and without the third conduction channel indicate that both models capture the experimental data. Since inclusion of the linear term does not appreciably change the fitted gap value, it is excluded from subsequent analysis. 
\begin{figure*}[!h]%
\centering
\includegraphics[width=0.5\textwidth]{SI_gap_closure.eps}
\caption{A comparison of the high field behavior of the gap extracted from an Arrhenius analysis (rising field only) and parallel conduction fits (rising and falling fields). All methods give similar gap behavior and demonstrate the gap closes non-monotonically at high fields.}\label{SI_gap_closure}
\end{figure*}

Having demonstrated that the gap closes non-linearly at high fields, we now confirm the results of our Arrhenius fit by comparing those findings to the high field dependence of the gap acquired from parallel conduction fits with the model describeed by Eq. (\ref{eq_conduction}). As shown in Figure \ref{SI_gap_closure}, the high field dependences of the gap acquired from the Arrhenius analysis and from the parallel conduction plots are in good agreement, even when the falling field data is analyzed. This provides additional evidence that the gap does not monotonically close at high fields.

\section{Gap Closure Along [011] Crystallographic Axis}

In addition to examining the field-induced gap closure for fields applied along the [100] crystallographic axis, similar measurements were performed with field applied along the [011] crystallographic axis. Figure \ref{SI_110_gap_closure}a shows the field-dependence of the magnetoresistance acquired at fixed temperature. This data possesses a similar field dependence to the [100] data shown in Fig. 1a of the main text. Quantum oscillations in the insulating state of YbB$_{12}$ are also present in this measurement configuration at high fields. An Arrhenius analysis (Fig. \ref{SI_110_gap_closure}b) indicates gap closure with the field applied along [011] is similar to gap closure with the field along [100]. Figure \ref{SI_110_gap_closure}c demonstrates this similarity by comparing the field-dependence of the gap when the field is applied parallel to [100] and [011] with fields normalized by the insulator-metal transition field. These data show that the gap closes non-linearly for fields applied along multiple crystallographic axes, consistent with the possibility of an excitonic insulator transition prior to the insulator-metal transition. Note, an Arrhenius analysis at lower field values for the [011] case was unreliable owing to an insufficient amount of high temperature data.

\begin{figure*}[!h]%
\centering
\includegraphics[width=1\textwidth]{SI_110_gap_closure.eps}
\caption{(a) YbB$_{12}$ magnetoresistance with the magnetic field applied along the [011] crystallographic axis. Along this direction the insulator-metal transition occurs at $\sim$54 T. (b) An Arrhenius analysis of the resistivity was used to obtain the evolution of the gap as a function of field in pulsed fields (PF). (c) The gap closure for fields along the [100] and [011] show similar deviations from linear gap closure when the field is normalized by the insulator-metal transition in each direction.}\label{SI_110_gap_closure}
\end{figure*}

\section{Checks for Field-Induced Heating Effects}

Field-induced heating in pulsed field experiments can arise from induced Eddy currents or magnetocaloric effects~\cite{Terashima2018}. Eddy current heating can be a substantial problem for metallic samples in pulsed fields, but can also be an issue in insulating samples owing to their metallic contacts.

\begin{figure*}[!h]%
\centering
\includegraphics[width=0.9\textwidth]{SI_eddy.eps}
\caption{Field-induced heating effects were assessed by using pulses with different maximum fields at the same temperature. The (a) rising and (b) falling field data for these pulses indicate minimal field-induced heating below $\sim35$T. There are small differences in (c) the rising field data above $\sim42$T and (d) the falling field data above $\sim35$T. These effects do not change the positions of the quantum oscillations.}\label{SI_eddy}
\end{figure*}

Eddy current heating typically scales with $H^2$, so we performed a series of pulses with different maximum fields to check for field-induced heating effects. As shown in Fig. \ref{SI_eddy}, the field dependence of the resistivity below $\sim35$T is independent of the maximum field. The rising field data possesses slight differences above $\sim42$T, as does the falling field data above $\sim35$T. However, directly comparing the rising and falling field data for the 44T and 52T pulses (Fig. \ref{SI_eddy_rising_falling}) shows the position of the quantum oscillation is not impacted by these effects. To further minimize the impact of these effects on our conclusions, where possible, we limited ourselves to analyzing data acquired while the sample was immersed in liquid and from pulses with similar maximum fields.

It is worth noting that there is a small, and systematic, difference between the positions of the quantum oscillations in the rising and falling field data even when the maximum field is chosen to keep the sample within the insulating state (e.g., Fig. \ref{SI_eddy_rising_falling}). Though this shift is small and does not affect our conclusions, we focused our analysis on the rising field data to minimize its impact. The origin of this observation is unclear, but could be caused by Eddy current heating of the electrodes or magnetocaloric effects~\cite{Terashima2018} in the sample close to the transition. Note, these effects are present even when the maximum field is kept below the insulator-metal transition, precluding effects stemming from entry into the metallic state (e.g., hysteretic effects).

\begin{figure*}[!h]%
\centering
\includegraphics[width=0.5\textwidth]{SI_eddy_rising_falling.eps}
\caption{There is a small, systematic shift between the positions of the quantum oscillations during the rising and falling fields. This shift occurs even when the sample does not enter the field-induced metallic state.}\label{SI_eddy_rising_falling}
\end{figure*}

\newpage
\section{Checks for Joule Heating Effects}

Joule heating effects were analyzed by varying the current used for resistivity measurements during a sequence of 60T pulses at similar temperatures and a fixed angle. As shown in Fig. \ref{SI_joule}, Joule heating in the insulating state causes the measured resistivity to shift to lower values for fixed magnetic fields. This trend is consistent with the temperature dependent magnetoresistance shown in Fig. 1 of the main text. These effects do not alter the position or shape of the quantum oscillations in an appreciable manner, however we utilized the smallest current possible to minimize impacts of Joule heating.

\begin{figure*}[!h]%
\centering
\includegraphics[width=1\textwidth]{SI_joule.eps}
\caption{Joule heating effects were assessed by varying the current used to measure the resistivity during the pulsed field experiments. For a given field, an increase in current caused a reduction in resistivity in both the (a) rising field and (b) falling field, consistent with Joule heating. Importantly, the positions and shapes of the quantum oscillation in both the (c) rising field and (d) falling field are minimally affected. Vertical dashed lines are guides for the eye.}\label{SI_joule}
\end{figure*}

 \newpage

\bibliography{sn-bibliography}
